\documentclass[twocolumn,aps,prd]{revtex4-2}
\AtBeginDocument{\renewcommand{\natexlab}[1]{#1}}
\usepackage{slashed}
\usepackage{enumitem}
\usepackage{tikz-cd}
	\usepackage{amsmath}
	\usepackage{amsfonts}
	\usepackage{amssymb}
	\usepackage{graphicx}
	\usepackage{mathtools}
	\usepackage{stmaryrd}
	\usepackage{autobreak} 
	\allowdisplaybreaks
	\usepackage[colorlinks=true,citecolor=blue,linkcolor=red]{hyperref}
	\usepackage{bbold}					
	\usepackage[makeroom]{cancel}		
	\usepackage{multirow}				
	\usepackage[normalem]{ulem}        
	\usepackage{array} 
\usepackage{lipsum}
\usepackage{graphicx}

\usepackage{titletoc} 
\usepackage{mathtools}
\usepackage{soul}

	\newcolumntype{x}[1]{>{\centering\let\newline\\\arraybackslash\hspace{0pt}}p{#1}}
	
	\makeatletter
\newcommand*\rel@kern[1]{\kern#1\dimexpr\macc@kerna}
\newcommand*\widebar[1]{%
  \begingroup
  \def\mathaccent##1##2{%
    \rel@kern{0.8}%
    \overline{\rel@kern{-0.8}\macc@nucleus\rel@kern{0.2}}%
    \rel@kern{-0.2}%
  }%
  \macc@depth\@ne
  \let\math@bgroup\@empty \let\math@egroup\macc@set@skewchar
  \mathsurround\z@ \frozen@everymath{\mathgroup\macc@group\relax}%
  \macc@set@skewchar\relax
  \let\mathaccentV\macc@nested@a
  \macc@nested@a\relax111{#1}%
  \endgroup
}
\makeatother

	\DeclareMathOperator{\sign}{sign}  	

	\DeclareMathAlphabet{\mathbbold}{U}{bbold}{m}{n}

	\def\PRLequal{\,{=}\,}
	
	\def\PRLequiv{\,{\equiv}\,}
	\def\PRLminus{\,{-}\,}
    \def\PRLpm{\,{\pm}\,}
	\def\PRLplus{\,{+}\,}
    \def\PRLin{\,{\in}\,}
    \def\PRLto{\,{\to}\,}

    \def\PRLto{\,{\to}\,}
    \def\PRLtimes{\,{\times}\,}
    \def\PRLcdot{\,{\cdot}\,}

	\newcounter{subeqn} %
	\makeatletter
	\@addtoreset{subeqn}{equation}
	\makeatother

\definecolor{XQ}{rgb}{1,0,0}
\definecolor{ZM}{rgb}{.5,0,.5}
\definecolor{SD}{rgb}{0,1,0}

\newcommand\trick[1]{} 

\begin{document}
\title{Mixed state topological order parameters for symmetry protected fermion matter}

\author{Ze-Min Huang$^{1,2}$}

\author{Sebastian Diehl$^1$}

\affiliation{$^1$Institute for Theoretical Physics, University of Cologne, 50937 Cologne, Germany}
\affiliation{$^2$Joint Quantum
Institute, University of Maryland, College Park, Maryland 20742, USA}

\begin{abstract}
We construct an observable mixed state topological order parameter for symmetry-protected free fermion matter. It resolves the entire table of topological insulators and superconductors, relying exclusively on the symmetry class, but not on unitary symmetries. It provides a robust, quantized signal not only for pure ground states, but also for mixed states in- or out of thermal equilibrium.  Key ingredient is a unitary probe operator, whose phase can be related to spectral asymmetry, in turn revealing the topological properties of the underlying state. This is demonstrated analytically in the continuum limit, and validated numerically on the lattice. The order parameter is experimentally accessible via either interferometry or full counting statistics, for example, in cold atom experiments. 

\end{abstract}
\maketitle

{\color{red}\textit{Introduction.--}} The integer quantum Hall effect  represents a paradigmatic instance of a topological phase \cite{girvin1986springer, stone1992world}. Its  topological properties are encapsulated in the Thouless-Kohmoto-Nightingale-den Nijs (TKNN) invariant for the ground-state wavefunction, directly measurable via the Hall conductance as a topological order parameter \cite{thouless1982prl, niu1985prb}.  The spectrum of fermion topological phases has since widened dramatically, enriched by the integration of symmetries \cite{altland1997prb,nayak2008rmp,ryu2010njp,kitaev2003aop, kitaev2009aip,  qi2011rmp, hassan2010rmp, chen2013prb,ludwig2016ps,chiu2016rmp}, resulting in a diverse array of topological invariants \cite{kitaev2001pu,volovik2003oxford, kane2005prlqsh, kane2005prlz2,levin2006prl, li2008prl, kitaev2006prl,fu2006prb,fu2007prb,sheng2006prl,moore2007prb, fu2007prb,roy2009prbz2, prodan2009prb,qi2010prb,pollmann2010prb,turner2011prb, qi2012prl, bulmash2015prx,kapustin2015jehp,chiu2016rmp, hassan2017prbEntanglement, hassan2017prl, shizoki2018prb}. These are defined through momentum- or phase-space topological invariants \cite{kitaev2001pu, kane2005prlqsh,volovik2003oxford, fu2006prb, kane2005prlz2,sheng2006prl,moore2007prb, fu2007prb,roy2009prbz2, prodan2009prb,qi2010prb, kapustin2015jehp, bulmash2015prx,chiu2016rmp}, twisted-boundary conditions and spacetime manifold surgery \cite{kapustin2015jehp, hassan2017prl,shizoki2018prb}, and information-theoretic quantities \cite{kitaev2006prl,levin2006prl, li2008prl, qi2012prl,turner2011prb, pollmann2010prb, hassan2017prbEntanglement}. 
Despite their mathematical elegance, these  frameworks often lack direct connection to experimental setups. Ongoing efforts to bridge this gap proceed through specific instances, such as detecting the polarization operator or the Berry phase in Rice-Mele-, Chern- or quantum spin Hall insulators~\cite{resta1998prl, grusdt2014pra, dwzhang2018aip, wawer2021prbz2,huang2022prb}. For  superconductors with integer valued invariant, fermion parity pumping has been explored \cite{ohyama2022prb}. A single unifying quantity encompassing the full periodic table, and equipped with a clear physical interpretation -- leading the way to observation -- is outstanding. 

In targeting such observables, rapid developments in engineered quantum systems, such as cold atomic gases \cite{bloch2012np,cirac2012np,goldman2016np} or more generally noisy intermediate-scale quantum (NISQ) platforms \cite{semeghini2021science, satzinger2021science, mi2023arxiv}, come to aid. They unleash novel preparation strategies in- and out-of-equilibrium, as well as diagnostic tools, including the detection of global observables. Yet, all of these platforms are inevitably exposed to various forms of noise and decoherence channels, so that their idealization by pure ground states of a Hamiltonian might be inappropriate. An additional desideratum of a physically accessible topological signal thus is its robustness against perturbations that degrade pure to mixed state. More ambitiously, it should be able to detect topological phase transitions in mixed state density matrices~\cite{viyuela2014prl1d, viyuela2014prl2d,huang2014prl, budich2015prb, altland2021prx, huang2022prb,sieberer2023arxiv}.

Here we construct such a universally applicable and robust observable. We show that it provides a mixed state topological order parameter across all classes of symmetry protected, free fermion matter: It takes quantized values and thus exhibits sharp boundaries between topologically distinct mixed states. The construction requires minimal input -- the symmetry class. It builds on relating a phase signal to the spectral asymmetry of a new object, the asymmetry matrix $Q_W$. The latter reduces to the Hamiltonian matrix for the prime series, characterized by integers in $\mathbb{Z}$ or $2\mathbb{Z}$, but more generally enables a finer resolution and exhausts the entire periodic table. The phase signal is observable in experiments with cold atoms, e.g. via interferometry \cite{sjoqvist2000prl, bardyn2018prx} or, as we elaborate, full counting statistics of global operators (such as total particle number or spin) \cite{levitov1993jetp,bakr2009nature, sherson2010nature, haller2015nature, anton2017science, humeniuk2019prl}.

{\color{red}\textit{A topological order parameter for all symmetry classes.--}} We define the order parameter via the phase of the expectation value of a unitary probe operator 
\begin{align}
\phi_{W}[w]&\PRLequal\text{arg}[\mathcal{Z}_W[w]], \ \ \mathcal{Z}_W[w]\PRLequal\langle e^{-i\hat{{W}}[w]} \rangle, \label{eq:solution}\\\nonumber
\hat{{W}}[w]&\PRLequal\sum_{i, a, b}\hat{\psi}^\dagger_{i, a}w_i\mathcal{W}_{ab} \hat{\psi}_{i, b} , 
\end{align}
where $\langle ... \rangle \PRLequal \mathrm{Tr} [\hat \rho ... ]/ \mathrm{Tr} [\hat \rho] $, $\hat{\psi}_{i, a} (\hat{\psi}^\dag_{i, a})$ are the fermionic annihilation (creation) operators at site $i$ with internal indices $a$. The state of the system is characterized by its density matrix $\hat{\rho}\PRLequal e^{-\hat{G}}$, taken to be Gaussian, i.e., $\hat{G}\PRLequal\sum_{ij, ab} \hat{\psi}^\dagger_{i,a}G_{ij,ab}\hat{\psi}_{j,b}$: $G$ is local in real space; the topological signal will be activated in the presence of background gauge fields, $G\PRLequal G[A]$. Such density matrix represents gapped fermionic matter in different symmetry classes in- \cite{ryu2010njp} and out-of-equilibrium \cite{diehl2011np, viyuela2014prl1d, viyuela2014prl2d,huang2014prl, bardyn2013njp, altland2021prx, huang2022prb}. Pure states obtain in zero temperature ground states ($\hat G \PRLequal \beta \hat H$ for inverse temperature $\beta $ and Hamiltonian $\hat H$, in the limit $\beta \PRLto \infty$), or as dark states of Lindbladian evolution \cite{altland2021prx, huang2022prb}. The probe operator is characterized by a possibly space dependent real function $w_i$, and a Hermitian matrix in internal space $\mathcal {W}$, with the constraint $\mathcal {W}^2 \PRLequal\mathbb{I}$. We will demonstrate that for a suitable choice of $\mathcal{W}$, determined exclusively by the symmetry class, the winding of $w$ will activate the topological charge contained in $\hat{G}$,  across the entire periodic table of topological insulators and superconductors. This winding features the product of momentum- and real-space topological invariants. The momentum space topological invariant includes examples like the Chern number. The real-space topology derives from the homotopic characteristics of the background gauge fields within $\hat{G}[A]$. The latter bifurcate into two classes, either $U(1)$ gauge fields for $U(1)$ symmetric insulators, or $\mathbb{Z}_2$ gauge fields for superconductors with fermion parity symmetry. Our focus is on the $U(1)$ case, with the necessary adjustments for $\mathbb{Z}_2$ provided subsequently.

We now present the recipe for constructing the function $w$ and the matrix $\mathcal{W}$, before proceeding to its derivation.
The choice of $w$ is determined exclusively by the spatial dimension:
\begin{enumerate}[label=(\Alph*)]
\item Even spatial dimensions: $w$ is a spatially homogeneous constant with value $[0,\, 2\pi]$. It spans a parameter cycle, such that $\frac{\Delta\phi_W}{2\pi} \PRLequiv\int_{0}^{2\pi}\frac{dw}{2\pi}\partial_{w}\phi_{W} \PRLin \{\mathbb{Z},\, 2\mathbb{Z},\, \mathbb{Z}_2\}$ renders a phase winding number, detecting the underlying topology.\label{list: a}
\item Odd spatial dimensions: $w\PRLequal w_i$ is an inhomogeneous function, varying slowly in one spatial direction and constant otherwise, respecting periodic boundary conditions (e.g., $\frac{2\pi\boldsymbol{x}_\alpha}{N_\alpha}$ with $\boldsymbol{x}_\alpha$  the coordinate and $N_\alpha$ the site number along the $\alpha$-th spatial dimension). The order parameter is then $\frac{\phi_W}{\pi}$ itself, i.e., $\frac{\phi_W}{\pi}\PRLin \{\mathbb{Z},\, 2\mathbb{Z},\, \mathbb{Z}_2\}$.
\label{list: b}
\end{enumerate}
These choices will be rationalized from a dimensional reduction scheme connecting even- to odd dimensions. 

\begin{figure}
\includegraphics[scale=0.3,angle=90]{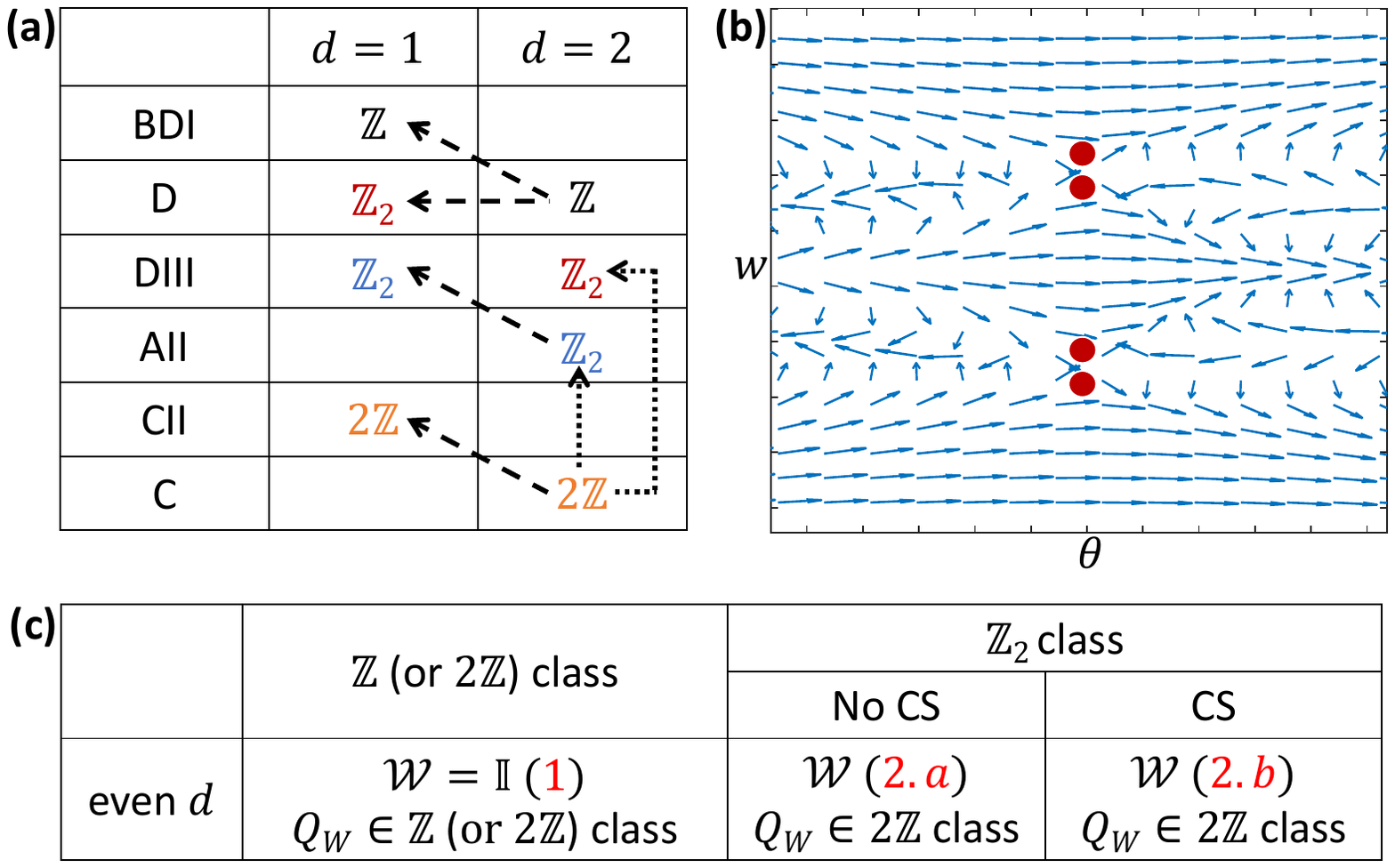}

\caption{Mixed state order parameter concept. (a)\,In even spatial dimensions ($d\PRLequal2$ here), we construct the mixed state order parameter by choosing the probe matrix  $\mathcal{W}$ (Eq.~\eqref{eq:solution}) such that $Q_W$ (Eq.~\eqref{eq:QW}) belongs to the $2\mathbb{Z}$ class, serving as a parent for $\mathbb{Z}_2$ classes (dotted arrows). Results in odd spatial dimensions ($d\PRLequal1$ here) are inferred via dimensional reduction (dashed arrows). (b)\,The reduction to $\mathbb{Z}_2$ arises from an ambiguity in the choice of $\mathcal{W}$: different $\mathcal{W}$ change $\frac{\Delta\phi_W}{2\pi}$ by $4\mathbb{Z}$, manifesting as the appearance of a quartic number of Fisher zeros in the parameter space for the complex amplitude $\mathcal{Z}_W[w]$ (Eq.~\eqref{eq:solution}), where $\theta$ parameterizes different $\mathcal{W}$. 
(c)\,Overview of the choices of $\mathcal{W}$.\label{fig:conceptual_plot}}
\end{figure}

The form of $\mathcal{W}$ is instead determined by symmetry. Symmetry protected fermion matter falls into two categories  \cite{ryu2010njp, chiu2016rmp, bardyn2013njp, altland2021prx}: (i)\,The prime series with integer topological invariant, $\mathbb{Z}$ or $2\mathbb{Z}$ valued;  
and (ii)\,the descendant states with $\mathbb{Z}_2$ invariant. The form of $\mathcal{W}$ changes accordingly (see
Fig.~\ref{fig:conceptual_plot} for an overview): 
\begin{enumerate}
\item For the prime series, $\mathcal{W}\PRLequal\mathbb{I}$.  \label{list:1}
\end{enumerate}
In this case, $\sum_{a,b, i}\hat{\psi}_{a, i}^\dagger\mathcal{W}_{ab}\hat{\psi}_{b, i} \PRLequal\hat Q$ is the charge operator. However, our scheme relies on the defining discrete symmetry of the class alone, and not on 
continuous unitary symmetry, like $U(1)$ or $SU(2)$ associated to charge and spin conservation;
$\hat G$ needs not to commute with $\hat Q$. 

For the descendant states, a symmetry constraint upon $\mathcal{W}$ is needed:
\begin{enumerate}[label=2.\alph*]
\item For systems without chiral symmetry (`No CS' in Fig.~\ref{fig:conceptual_plot}\,(c)), we require that $\mathcal{S}\mathcal{W}\mathcal{S}^{-1}\PRLequal \PRLminus\mathcal{W}$, with $\mathcal{S}$
the matrix representation of the protecting time-reversal $\left(\mathcal{T}\right)$,
or particle-hole symmetry ($\mathcal{C}$) \cite{fn_ph}. \label{list:2a}

\item For systems with chiral symmetry (`CS' in Fig.~\ref{fig:conceptual_plot}\,(c)), we require that (a)\,$\mathcal{W}$ anti-commute with chiral
symmetry, (b)\,$\mathcal{T}\mathcal{W}\mathcal{T}^{-1}\PRLequal\PRLminus \mathcal{W}$ (or $\mathcal{C}\mathcal{W}\mathcal{C}^{-1}\PRLequal\PRLminus\mathcal{W}$)
in $4k\PRLplus 2$ (or $4k$) spatial dimensions,
with $k\in\mathbb{N}_0$. \label{list:2b} 
\end{enumerate}
 This will be explained by the $\mathcal{W}$ matrix being inferred from a $2\mathbb{Z}$ class parent in the same dimension (see Fig.~\ref{fig:conceptual_plot}\,(a)). For a fixed $\mathcal{W}$, the ensuing accumulated phase signal consequently satisfies $\frac{\Delta\phi_W}{2\pi}\PRLin 2\mathbb{Z}$. The reduction to $\mathbb{Z}_2$ then arises from the freedom in choosing $\mathcal{W}$ matrices compatible with the above symmetries, demonstrating that the signal is defined only modulo $4\mathbb{Z}$.  We take advantage of this mechanism to detect the $\mathbb{Z}_2$ signal $\frac{\Delta\phi_W}{2\pi}$ via random sampling of $\mathcal{W}$. 
 
In this discussion, we have assumed even dimension. For the odd dimensional cases, the $\mathcal{W}$ matrix can be taken the same as for its even dimensional parent state, following from dimensional reduction \cite{qi2010prb, huang2022prb, FN_od} (see Fig.~\ref{fig:conceptual_plot}\,(a)).
We now derive these results, and illustrate them via examples. To this end, we will pass to the continuum limit, where 
$G_{ij,ab}$ becomes a first-quantized local operator, and $w_i$ a function on $d$-dimensional space; our results are expected to equally hold on the lattice  \cite{knill2017arxiv}, as confirmed by our numerical simulations.

{\color{red}\textit{$\Delta\phi_W$ measures activated topological charge.--}}
We focus on Eq.~\eqref{eq:solution}, and demonstrate that the winding of $w$ activates quantized topological charge of $\hat G$, measured by $\Delta\phi_W$. This will be achieved by identifying $\frac{\Delta\phi_W}{2\pi}$ as a \textit{spectral asymmetry}, which is a topological index \cite{atiyah1975cup, atiyah1976cup}.  
More explicitly, upon tracing out fermion degree of freedom in Eq.~\eqref{eq:solution}, one obtains 
\begin{eqnarray}
\phi_{W}&\PRLequal&\text{Im}\ln\det\{\cos\frac{w}{2}+i\sin(\frac{w}{2}) [\mathcal{W}\tanh\left(\frac{G}{2}\right)]\}.
\label{eq:phi_W_trace}
\end{eqnarray}
It represents a multi-valued function due to the presence of a branch cut in the logarithm \cite{wtrace_fn}. Specifically, upon performing a winding $w\PRLto w \PRLplus 2\pi$, this multi-valuedness singles out the topological charge encapsulated in $G$, manifesting itself as the spectral asymmetry of the following Hermitian \textit{asymmetry matrix} $Q_W$ \cite{supp},
\begin{equation}\label{eq:QW}
Q_W\equiv \frac{1}{2}\{ \mathcal{W}, \text{sign}(G)\}
\end{equation}
with $\text{sign}(G)\PRLequiv \frac{\tanh(\frac{G}{2})}{\sqrt{\tanh(\frac{G}{2})^2}}$ for gapped $G$, 
and 
\begin{equation}
\frac{\Delta\phi_W}{2\pi} 
\PRLequal \frac{1}{2}\sum_n \text{sign}(q_{n}),\label{eq:spec_asym} 
\end{equation} 
with $q_{n}$ the eigenvalues of $Q_W$ \cite{volovik2003oxford}. Here the winding is implemented via spatially homogeneous $w$, the discussion for slowly varying $w$ is postponed to later. 
Clearly, for $\mathcal{W} \PRLequal \mathbb{1}$, $\frac{\Delta \phi_W}{2\pi}$ coincides with the spectral asymmetry of the operator $G$ itself, which will classify the prime series (see \ref{list:1}). Generally, other choices of $\mathcal{W}$  are necessary to resolve all symmetry classes. This is one of our key results: it identifies $Q_W$, rather than $G$, as the fundamental object to classify the topology of the entire table of symmetry protected fermion matter. 

According to Atiyah, Patodi and Singer~\cite{atiyah1975cup, atiyah1976cup}, the spectral asymmetry of an elliptic (local) Hermitian operator is related to topology. Here, this 
manifests as the product of momentum- (denoted as $\text{ch}_W$) and real-space ($\int \mathfrak{C}$) topological invariants,
\begin{equation}
\frac{\Delta\phi_W}{2\pi} \PRLequal\text{ch}_W \times \int d^d \boldsymbol{x} \, \mathfrak{C}(\boldsymbol{x})\in \mathbb{Z}, \label{eq:delta_phiW}
\end{equation}
which is the fundamental formula for our mixed state order parameter.
Specifically, $\text{ch}_W$ captures the topological properties of $Q_W$, characterized by the homotopy group, e.g.\,$\pi_{d+1}[GL(N, \mathbb{C})]$
for the map from frequency-momentum space (with dimension $d\PRLplus1$) to $\frac{1}{i\omega - Q_W}\PRLin GL(N, \mathbb{C})$. Clearly, for $\mathcal{W}\PRLequal\mathbb{I}$, this reproduces the Chern number in the Chern insulator.
Meanwhile, $\int \mathfrak{C}$ depends only on homotopic properties of the background field, with $\mathfrak{C}$ for the associated topological charge density \cite{niemi1984prd}. For the case of $U(1)$ gauge fields $A$ and in even spatial dimensions $d\PRLequal2n$, $\int \mathfrak{C}$ can be the background topological charge of a magnetic field, or the winding number of skyrmions \cite{wilczeck1983prl}, 
\begin{equation}
\mathfrak{C}^{(2n)}(\boldsymbol{x})\PRLequal \frac{\epsilon^{0i_1i_2\dots i_{2n-1} i_{2n}} }{(2\pi)^n n!}\partial_{i_1}A_{i_2}\dots\partial_{i_{2n-1}}A_{i_{2n}}.\label{eq:CW_2n}
\end{equation} 
Together, $\frac{\Delta\phi_W}{2\pi}$ captures the topological charge localized within real-space solitons \cite{goldstone1981prl, niemi1986pr}, underpinning its role as an order parameter of mixed state topology.

The spectral asymmetry $\frac{\Delta\phi_W}{2\pi}$ closely links to the fermion parity expectation value, but crucially provides a finer resolution. Specifically, the bridging formula is
\begin{align}
\text{sign}\langle(-1)^{\hat{Q}}\rangle\PRLequal e^{i \pi \times (\frac{\Delta \phi_W}{2\pi})},\label{eq_fermion_parity}
\end{align} 
established by recognizing that $e^{i\phi_W}|_{w\PRLequal\pi}$ coincides with the fermion parity, as a consequence of $\mathcal{W}^2\PRLequal\mathbb{I}$. Meanwhile, Hermiticity of $\mathcal W$ implies that $\phi_W(w)\PRLequal-\phi_W(-w)$, and thus $e^{i\phi_W}|_{w\PRLequal\pi}$ as well as the fermion parity discern only even-/oddness of $\frac{\Delta\phi_W}{2\pi}$, yielding a $\mathbb{Z}_2$ signature. The $\mathbb{Z}$-valued spectral asymmetry $\frac{\Delta\phi_W}{2\pi}$ instead captures a richer topological pattern.

Changes in $\frac{\Delta\phi_W}{2\pi}$ can only occur when $Q_W$ contains zero modes. This implies the appearance of Fisher zeros in $\mathcal{Z}_W$, which can be viewed as a non-unitary partition function. These fall into two categories. Indeed, the existence of a Fisher zero is a sufficient, but not a necessary condition for a topological phase transition: The latter occurs for zero modes of $G$ (which implies zero modes of $Q_W$). But zero modes of $Q_W$ can also occur while $G$ remains gapped, depending on the choice of $\mathcal{W}$ matrices. Symmetry (\ref{list:2a} and \ref{list:2b}) imposes crucial constraints, ensuring zeros of this type to appear in multiples of four (see Fig.~\ref{fig:conceptual_plot}\,(b)) \cite{supp}. This results in a $\mathbb{Z}_2$ classification descending from the $2\mathbb{Z}$-valued $\Delta\phi_W$, as we explain next.

{\color{red}\textit{Exhausting all symmetry classes by choice of $\mathcal{W}$.--}} We now construct $\mathcal{W}$ for all symmetry classes, and thus derive the conditions presented in \ref{list:1}, \ref{list:2a} and \ref{list:2b} based on the spectral asymmetry formula Eq.~\eqref{eq:delta_phiW}. The real-space topology $\mathfrak{C}$ stems from a $U(1)$ gauge field (see Eq.~\eqref{eq:CW_2n}) for insulators (represented by complex fermions, (i) and (ii) below), and $\mathbb{Z}_2$ gauge fields for superconductors (represented by Majorana fermions, (iii)).  

\textit{(i).\,Insulators in even spatial dimensions.--}  
For the $\mathbb{Z}$ and $2\mathbb{Z}$ classes,
the choice is simple: we take $\mathcal{W}\PRLequal \mathbb{I}$ (cf. \ref{list:1}), yielding $\frac{\Delta\phi_W}{2\pi}\PRLin \mathbb{Z}\ (\text{or}\ 2\mathbb{Z})$ as mixed state order parameter. This is possible since 
these classes possess a non-trivial homotopic invariant $\pi_{d+1}[GL(N, \mathbb{C})]$ \citep{volovik2003oxford,  qi2008prb,huang2022prb}; the $2\mathbb{Z}$ classification results from the presence of time-reversal/particle-hole symmetry requiring the appearance of an even number of $\mathbb{Z}$-class copies \cite{ryu2010njp}. 

The $\mathbb{Z}_{2}$ classes however, generally render a vanishing signal $\frac{\Delta\phi_{W}}{2\pi}$ when $\mathcal{W}\PRLequal\mathbb{I}$ (and thus positive fermion parity, cf. Eq.~\eqref{eq_fermion_parity}). 
This is because symmetry enforces the spectrum of $\sign(G)$ to be composed of 
symmetric pairs of eigenvalues with opposite sign in Eq.~\eqref{eq:spec_asym}~\cite{supp}.
We thus opt for a different choice of the $\mathcal{W}$ matrix, as listed in \ref{list:2a} and \ref{list:2b}, which endows an opposite sign to the eigenvalues of the symmetric pairs of $\sign(G)$. Namely, we first observe that the associated $Q_W$ is in the $2\mathbb{Z}$ class: Its spectral asymmetry takes even integer values $\frac{\Delta\phi_W}{2\pi}\PRLin 2\mathbb{Z}$, ensured by the positive fermion parity noted above. The reduction to $\mathbb{Z}_2$ then roots in the remaining freedom for choosing $\mathcal{W}$ (see \cite{supp} for a derivation): Varying $\mathcal{W}$ alters $\frac{\Delta\phi_W}{2\pi}$ by $4\mathbb{Z}$, manifesting in $\mathcal{Z}_W$ as a multiple of four for the number of Fisher zeros in the parameter plane, see Fig.~\ref{fig:conceptual_plot}\,(b). Hence, we identify $\mathbb{Z}_2 \PRLequal 2\mathbb{Z}\mod4$, and obtain $ \frac{\Delta \phi_W}{2\pi} \PRLequal0,\,2\mod 4$ as a mixed state order parameter.

As a simple illustration, consider the $0$-dimensional $\mathbb{Z}_2$ class without chiral symmetry (class D), with representative model, $G\PRLequal m\sigma^{z}$  \citep{ryu2010njp}, and $\sigma^{x, y, z}$ ($\sigma^0$) for Pauli (unit) matrices. The spectrum is symmetric about the origin, i.e., $\frac{\Delta\phi_W}{2\pi}\PRLequal0$ for $\mathcal{W}\PRLequal\mathbb{I}$ (Eq.~\eqref{eq:spec_asym}), due to the defining particle-hole symmetry:
$\mathcal{C}\PRLequal\sigma^{x}\mathcal{K}$, with $\mathcal{K}$ for complex conjugation. The only $\mathcal{W}$ satisfying \ref{list:2a}
are $\mathcal{W}\PRLequal\PRLpm\sigma^{z}$, which imparts an opposite sign within symmetric pairs of $G$:  $Q_W\PRLequal\PRLpm \sign(m)\sigma^0$,  possessing non-zero $\frac{\Delta \phi_W}{2\pi} \PRLequal \frac{1}{2}[\PRLpm 2\, \text{sign}(m) \mod 4]$  (the half-integer prefactor here is a typical artifact of the Dirac nature of this toy model, and can be made integer by a proper regularization \cite{ryu2010njp}). 

\textit{(ii).\,Insulators in odd spatial dimensions.--} For the order parameters in odd spatial dimension, a caveat associated with Eq.~\eqref{eq:delta_phiW} is that $\text{ch}_W$ generally vanishes, due to Bott periodicity (see, e.g., \cite{stone2011jpamt}). Still, one can access the underlying topology by invoking slowly varying $w(\boldsymbol{x})$ (cf. \ref{list: b}), and then establish an order parameter via dimensional reduction. 
To this end, we start from a generalization of $\phi_W$ in Eq.~\eqref{eq:phi_W_trace} to such $w(\boldsymbol{x})$ (see \cite{supp} for a derivation from a Dirac model)
\begin{equation}
\phi_W\PRLequiv \text{ch}_W\PRLtimes \int d^{2n}\boldsymbol{x}\mathcal{I}_W[w(\boldsymbol{x})] \PRLtimes \mathfrak{C}^{(2n)}(\boldsymbol{x}),\label{eq:effective_action}
\end{equation}
where $\mathfrak{C}^{(2n)}$ is given in Eq.~\eqref{eq:CW_2n}.
$\mathcal{I}_W[w(\boldsymbol x)]$ is a model-dependent multi-valued function with the key property $\mathcal{I}_{W}|_{w}^{w+2\pi}\PRLequal2\pi$, so as to uphold the spectral asymmetry (Eq.~\eqref{eq:delta_phiW}), i.e., $\frac{\phi_W|_{w\PRLequal0}^{w\PRLequal2\pi}}{2\pi}\PRLin \mathbb{Z}$.
We can then apply the standard dimensional reduction method \cite{qi2008prb, ryu2010njp} to infer the descendant formula, via integrating out one spatial dimension. The ensuing odd dimensional signal, 
with $\mathcal{W}$ read off from its parent state and inhomogenous $w$ (see Fig.~\ref{fig:conceptual_plot}\,(a) and \ref{list: b}), is \cite{supp}
\begin{equation}
\frac{\phi_W}{\pi} \PRLequal\text{ch}_W\times(\oint d \boldsymbol{x}_\alpha \frac{\partial}{\partial \boldsymbol{x}_\alpha}\mathcal{I}_W) \times [\frac{1}{2\pi}\int d^{2n-2} \boldsymbol{x}\mathfrak{C}^{(2n-2)}] \in \mathbb{Z},\label{eq:phi_odd}
\end{equation}
which is integer quantized when taking the background field in $\mathfrak{C}^{(2n-2)}$ to be independent of the extra coordinate $\boldsymbol{x}_\alpha$. This establishes $\frac{\phi_W}{\pi}$ as an order parameter in \ref{list: b}.

\textit{(iii).\,Superconductors.--} 
Thus far, our focus has been on insulators, i.e. systems with a conserved particle number. This allows us to introduce a $U(1)$ gauge field as a means to activate the spectral asymmetry. While a continuous symmetry is crucial for constructing $U(1)$ gauge fields, it is irrelevant for the definition of symmetry classes \cite{altland1997prb, schnyder2008prb}. Indeed, in the case of particle-hole symmetric fermion matter -- physically representing superconductors -- the continuous $U(1)$ symmetry is reduced to a discrete $\mathbb{Z}_2$ fermion parity symmetry. The central concept, spectral asymmetry, remains applicable, but is leveraged in a different manner: It is activated by a $\mathbb{Z}_2$ fermion parity symmetry gauge field instead, implemented through twisted spatial boundary conditions~\cite{karch2019scipost}. Hence, our recipe in \ref{list:1}, \ref{list:2a} and \ref{list:2b} generalizes to superconductors. We relegate the details to \cite{supp}, but illustrate its working in Example II below.

\begin{figure}
\includegraphics[scale=0.3, angle=90]{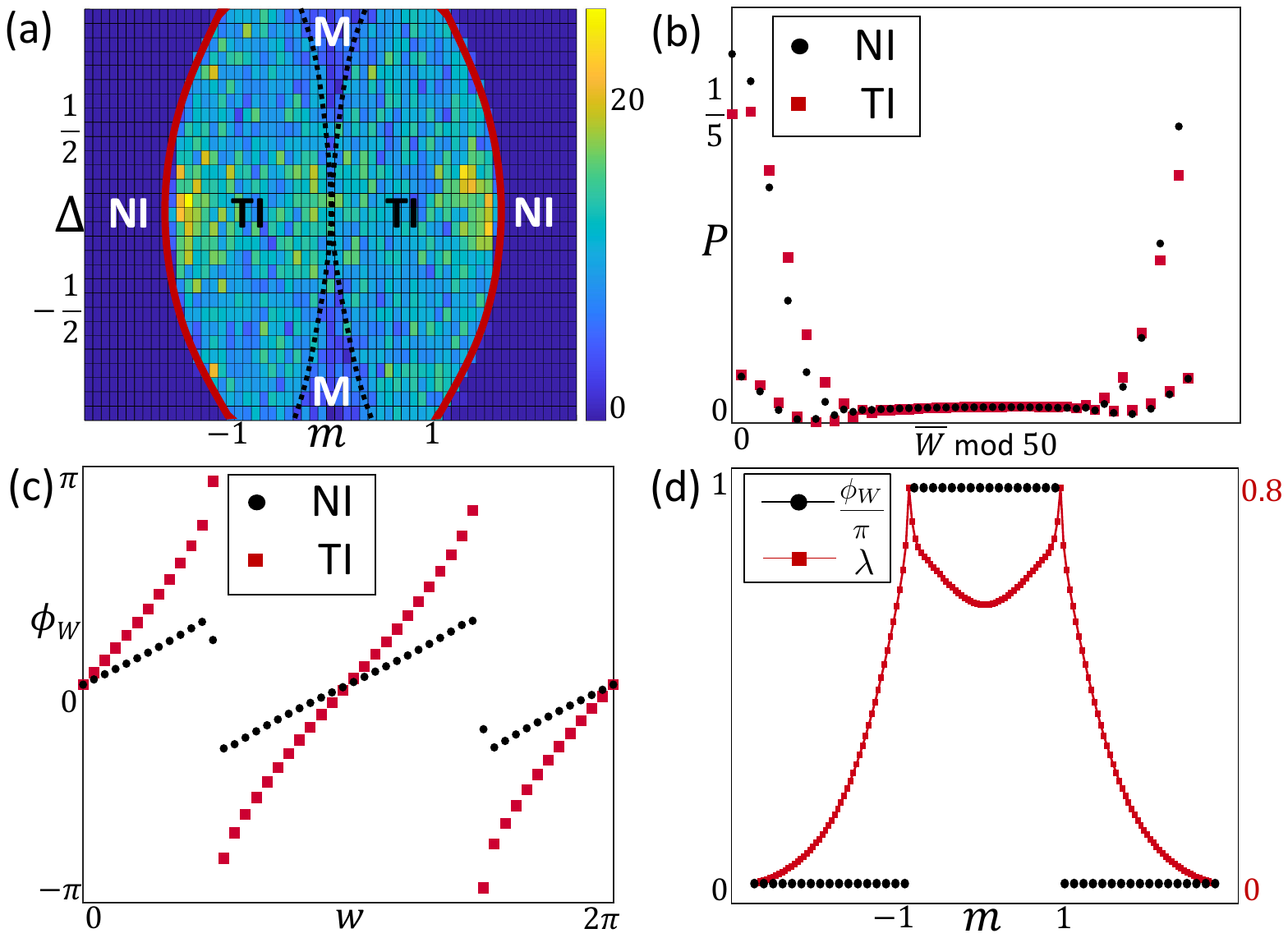}

\caption{Numerical results for the modified BHZ (mBHZ) model, (a)-(c), and the DIII superconductor, (d). (a)\,$\frac{\Delta\phi_W}{2\pi}$ as a signature for the topological (normal) phase in the mBHZ model under insertion of one magnetic flux quantum at temperature $1$ and site number $15\PRLtimes 15$. $\mathcal{W}$ is randomly sampled $50$ times. The color bar is a histogram counting the number of non-trivial valued $\frac{\Delta\phi_W}{2\pi}$, i.e., $\frac{\Delta\phi_W}{2\pi}\PRLequal2 \mod 4$ \cite{fig2_fn}.
TI, NI,  and M stand for topological, normal, and gapless phase, respectively. (b)\,Full counting statistics for the mBHZ model in the canonical ensemble at $\beta\PRLequal3$, with fixed $\mathcal{W}\PRLequal\sigma^z \otimes \tau^0$. Its Fourier components encode the mixed state order parameter $\phi_W$, which are plotted in (c), as a function of $w$, leading to a nontrivial value $\frac{\Delta\phi_W}{2\pi} \PRLequal2\PRLin 2\mathbb{Z} \mod 4$ in TI. (d)\,Mixed state topological order parameter for a DIII superconductor as a function of $m$, i.e.,\,$\frac{\phi_W}{\pi}\PRLequal\frac{1}{2}\PRLtimes(0,\,2\mod4)$, where $w\PRLequal\frac{2\pi \boldsymbol{x}_1}{N_1}$ and $\frac{1}{2}$ from the Majorana nature. The amplitude $\lambda\PRLequiv \PRLminus\frac{\ln |\mathcal{Z}_W|^2}{N_1}$ exhibits a cusp at the transition points.
\label{fig:numerical_BHZ}
}

\end{figure}
In the following we will illustrate our findings with concrete examples, concentrating on cases with $\mathbb{Z}_2$ classification. The $\mathbb{Z}_2$ signal can be distilled by randomly and repeatedly choosing the probe matrix $\mathcal{W}$, while respecting the symmetry constraints. For clarity, we will present results for equilibrium states $G\PRLequal \beta H$ only; non-equilibrium states including dynamical scenarios are discussed in \cite{supp}.

{\color{red}\textit{Example I: Two dimensional AII class.--}} We focus on the (modified) Bernevig-Hughes-Zhang model (mBHZ) \cite{bernevig2006science}, 
\begin{equation}\label{eq:mbhz}
H\PRLequal\left(\begin{array}{cc}
H_{0}\left(\boldsymbol{k}\right) & -i\Delta\tau^{y}\\
i\Delta\tau^{y} & H_{0}^{*}\left(-\boldsymbol{k}\right)
\end{array}\right), \ \mathcal{T}\PRLequal\sigma^{y}\otimes\tau^{0}\mathcal{K},
\end{equation}
with $\tau^{x, y, z}$ ($\tau^{0}$) for Pauli (unit) matrices, and $\boldsymbol{k}$ for momentum. $H_{0}\PRLequal\boldsymbol{d}\PRLcdot\boldsymbol{\tau}$ with $\boldsymbol{d}\PRLequal\left(\sin\left(k_{x}\right),\, \sin\left(k_{y}\right),\, m\PRLplus\cos\left(k_{x}\right)\PRLplus\cos\left(k_{y}\right)\right)$, and the $\Delta$ term is introduced to break the $z$-axis spin rotational symmetry. 

Possible choices for the probe matrix compliant with \ref{list:2a} are then $\mathcal{W}\PRLequal\{\boldsymbol{\sigma}\otimes\tau^{x,\ z,\ 0},\, \sigma^{0}\otimes\tau^{y}\}$. 
Numerical results are presented in Fig.~\ref{fig:numerical_BHZ}.
The mixed state topological order parameter accurately reconstructs  the zero temperature phase boundary (red line) from a finite temperature situation ($\beta \PRLequal1$ here).

The topological order parameter can be observed in cold atom experiments by taking simultaneous snapshots of all particles \cite{bakr2009nature, sherson2010nature,  haller2015nature, anton2017science}, to build the full counting statistics (FCS) of the global operator $\hat W$. The FCS signal is the distribution function for the eigenvalues $\overline{W}$ of the global operator $\hat{W}[w\PRLequiv 1]$ with $w$ taken to be constant $1$, represented as
\begin{align*}
    P[\overline{W}]\equiv 
    \langle\delta(\hat W[1]-\overline{W})\rangle \PRLequal  \int_0^{2\pi} dw e^{i w\overline{W}}\langle e^{-i  w\hat W[1]}\rangle.
\end{align*} 
Its Fourier components at frequency $w$ render the mixed state order parameter $\mathcal{Z}_W[w]$.
Results for representative points in the normal (NI) and topological (TI) phases are shown in
Fig.~\ref{fig:numerical_BHZ}\,(b,\,c): While the FCS histogram looks qualitatively similar in both phases, the topologically non-trivial character is clearly visible in TI, indicated by the winding number along tuning $w$.
Alternatively, Mach-Zehnder interferometry could be used \cite{sjoqvist2000prl,bardyn2018prx, wawer2021prbz2}, where $\hat W$ acts as the Hamiltonian for an adiabatically imprinted Loschmidt echo \cite{zvyagin2016ltp,heyl2018rpp, bhattacharya2017prb, heyl2017prb}. This requires immersing the atoms in a cavity, and engineering a probe Hamiltonian defined with $w$ and $\hat{W}$.

{\color{red}\textit{Example II: One dimensional DIII class.-- }} To illustrate the dimensional reduction method and the real (Majorana) fermion case, we consider the
DIII class superconductor in one dimension. Its parent class is AII
in two dimensions. Following the dimensional reduction program, we work with a spatially dependent function $w\PRLequal\frac{2\pi}{N_{1}}\boldsymbol{x}_{1}$ (cf.\,\ref{list: b}). From Eq.~\eqref{eq:phi_odd} we then find  $\frac{\phi_W}{\pi}\PRLequal\frac{1}{2}\text{ch}_{W}\PRLin \mathbb{Z}$,
with prefactor $\frac{1}{2}$ due to the Majorana nature, and $\text{ch}_W\PRLin 2\mathbb{Z}$ (Eq.~\eqref{eq:phi_odd}) inherited from its parent state. Numerical results are shown in Fig.~\ref{fig:numerical_BHZ}\,(d) for the following model,
\begin{equation}
H\PRLequal-\left(\sin k_{x}\sigma^{z}\right)\otimes\tau^{x}+\left(m+\cos k_{x}\right)\sigma^{0}\otimes\tau^{z},
\end{equation}
with particle-hole symmetry $\mathcal{C}\PRLequal\sigma^{0}\otimes\tau^{x}\mathcal{K}$, and time-reversal symmetry $ \mathcal{T}\PRLequal\sigma^{y}\otimes\tau^{0}\mathcal{K}$,
where the Nambu basis is $\hat{\Psi}_{k}\PRLequal(\hat{\psi}_{k},\,-i\sigma^{y}\hat{\psi}_{k}^{\dagger})$. According to \ref{list:2a}, we choose time-reversal odd probe matrices, a property realized by the Pauli matrices, i.e., $\mathcal{W} \PRLequal \boldsymbol{n}\PRLcdot\boldsymbol{\sigma}$ (or $\frac{1}{2}\boldsymbol{n}\PRLcdot\boldsymbol{\sigma}\otimes \tau^0$ in Nambu space) for a spin pointing in direction  
$\boldsymbol{n}\PRLequal\left(\sin\theta\cos\phi,\, \sin\theta\sin\phi,\, \cos\theta\right)$, and sample the angles randomly.
Fig.~\ref{fig:numerical_BHZ}\,(d) verifies our order parameter as a probe of the underlying topology.

{\color{red}\textit{Conclusion.-- }} We have developed an observable mixed state topological order parameter for symmetry protected fermion matter, which takes the defining symmetry as the only input.
At the heart of the possibility of exhausting the full periodic table lies a new descendant relation between the $2\mathbb{Z}$ and  $\mathbb{Z}_2$ classes in the same dimension. While concentrating on free fermions here, the order parameter remains well defined (with the same bilinear $\hat W$) for interacting fermion systems. Their mixed state physics might be enriched by the activation of defects and novel entropy driven topological phase transitions. Exploring these effects  
will pave the way towards a thorough understanding of mixed state topology. A further direction is to connect the symmetry protected topology of mixed fermion states to topological order and the robustness of quantum information \cite{dennis2002jmp, Hastings2011, 
bao2023arxiv,  lee2022arxiv}. 

\section*{Acknowledgment}
The authors wish to thank Florian Meinert and Xiao-Qi Sun for discussions, and Lutian Zhao for his insights regarding the index theorem on a lattice. Z.-M.H. acknowledges the support from the JQI postdoctoral fellowship at the University of Maryland. S.D. is supported by the  Deutsche Forschungsgemeinschaft (DFG, German Research Foundation) under Germany’s Excellence Strategy Cluster of Excellence Matter and Light for Quantum Computing (ML4Q) EXC 2004/1 390534769 and by the DFG Collaborative Research Center (CRC) 183 Project No. 277101999 - project B02.

\clearpage 
\onecolumngrid

\begin{center}
\textbf{\large Supplemental Material for "Mixed state topological order parameters for symmetry protected fermion matter"}
\end{center}

\setcounter{equation}{0}
\setcounter{figure}{0}
\setcounter{table}{0}
\setcounter{page}{1}
\setcounter{section}{0}
\makeatletter
\renewcommand{\theequation}{S\arabic{equation}}
\renewcommand{\thefigure}{S\arabic{figure}}
\renewcommand{\thesection}{S\arabic{section}}
\renewcommand{\bibnumfmt}[1]{[S#1]}
\renewcommand{\citenumfont}[1]{S#1}

This supplemental material includes details for: (i) $\frac{\Delta\phi_W}{2\pi}$ as the spectral asymmetry for $Q_W$; (ii) Derivation of the effective action (Eq.~\eqref{eq:effective_action} in the main text) from a microscopic Dirac model; (iii) How the reduction from $2\mathbb{Z}$ to $\mathbb{Z}_2$ arises via the freedom of choosing $\mathcal{W}$; (iv) Details of the dimensional reduction from even to odd space dimensions; (v) Numerical results for non-equilibrium mixed state evolution; (vi) Numerical results for $\phi_W$ in cold atomic ensembles;
and (vii) Details on the mixed state order parameter construction for superconductors represented by Majorana fermions.


\tableofcontents{}

\section{$\frac{\Delta\phi_{W}}{2\pi}$ can change only when $Q_W$ is gapless \label{supp:sec_spec_asym}}

We demonstrate that for a fermionic Gaussian state
$\hat{\rho}\PRLequal e^{-\hat{\psi}^{\dagger}G\hat{\psi}}$, $\frac{\Delta\phi_{W}}{2\pi}$
depends only on the spectral asymmetry of $Q_W$, providing the details for the derivation of Eq. (\ref{eq:phi_W_trace}).

\subsection{Tracing out fermion degree of freedom}

We first trace out fermion degree of freedom
to reproduce Eq. (\ref{eq:phi_W_trace}) in the main text. For concreteness,
we shall assume $\hat{\psi}$ to be complex fermion, while derivations
for the Majorana case (or fermions in the Nambu space) is parallel, as we shall
comment along the way. The object of interest is 
\begin{equation}
\phi_{W}\left(w\right)\equiv\arg\left[\frac{1}{\text{Tr}\hat{\rho}}\text{Tr}\left(\hat{\rho}e^{-iw\hat{\psi}^{\dagger}\mathcal{W}\hat{\psi}}\right)\right],\ \hat{\rho}\PRLequal e^{-\hat{\psi}^{\dagger}G\hat{\psi}},
\end{equation}
which after tracing out fermions, yields 
\begin{eqnarray}\label{eq:S2}
\phi_{W}\left(w\right) & \PRLequal & \text{arg}\left[\frac{\det\left(\mathbb{I}+e^{-G}e^{-iw\mathcal{W}}\right)}{\det\left(1+e^{-G}\right)}\right]\nonumber \\
 & \PRLequal & \text{arg}\left[\det\left(e^{-i\frac{w}{2}\mathcal{W}}\right)\frac{\det\left(e^{i\frac{w}{2}\mathcal{W}}+e^{-i\frac{w}{2}\mathcal{W}}e^{-G}\right)}{\det\left(1+e^{-G}\right)}\right]\nonumber \\
 & \PRLequal & -\frac{w}{2}\text{tr}\mathcal{W}+\text{arg}\left\{ \det\left[\cos\left(\frac{w}{2}\right)+i\sin\frac{w}{2}\mathcal{W}\tanh\left(\frac{G}{2}\right)\right]\right\} .
\end{eqnarray}
Here, we have used the following identity for complex fermion
\begin{equation}
\text{Tr}\left(e^{-\hat{\psi}^{\dagger}A\hat{\psi}}e^{-\hat{\psi}^{\dagger}B\hat{\psi}}\right)\PRLequal\det\left(\mathbb{I}+e^{-A}e^{-B}\right),
\end{equation}
while its counterpart for Majorana fermions $\hat{\gamma}$ is  \cite{klich2014jsm}
\begin{equation}
[\text{Tr}\left(e^{-\hat{\gamma}A\hat{\gamma}}e^{-\hat{\gamma}B\hat{\gamma}}\right)]^2\PRLequal\det(\mathbb{I}+e^{-2A}e^{-2B}).
\end{equation}

\subsection{$\frac{\Delta\phi_{W}}{2\pi}$ as the spectral asymmetry of
$Q_W$}

Now we demonstrate one of our key results, that $\frac{\Delta\phi_{W}}{2\pi}$
only depends on the spectral asymmetry of the Hermitian matrix 
\begin{equation}\label{eq:QWdef}
    Q_W\PRLequal\frac{1}{2}\{\mathcal{W}, \frac{\tanh{(\frac{G}{2}})}{\sqrt{\tanh^2(\frac{G}{2}})}\}.
\end{equation}
Hence, $\Delta\phi_{W}$ can change only when the gap of $Q_W$ closes.

The proof is based on the following two observations: \\
1. Due to Hermiticity of $\mathcal{W}$:
\begin{align}\label{eq:hermiticity}
  \phi_{W}\left(w\right)\PRLequal-\phi_{W}\left(-w\right)  \implies \Delta\phi_{W}\PRLequal2\int_{0}^{\pi}dw\partial_{w}\phi_{W}\left(w\right) = 2 \phi(\pi),
\end{align}
which holds since $\phi(0) = 0$ (cf. Eq.~\eqref{eq:S2}).\\
2. $\phi_{W}\left(w=\pi\right)$ depends solely on the spectral
asymmetry of $Q_W$. 

To demonstrate 2., we first use observation 1. and the explicit form of $\phi(\pi)$ (Eq.~\eqref{eq:S2}) to get   
\begin{equation}
\Delta\phi_{W}=2\text{Im}\text{Tr}\ln\left[i\mathcal{W}\tanh\left(\frac{G}{2}\right)\right]
  =  2\text{Im}\text{Tr}\ln \left[i\mathcal{W}\frac{\tanh({\frac{G}{2}})}{\sqrt{\tanh^2(\frac{G}{2})}}\times\sqrt{\tanh^{2}\left(\frac{G}{2}\right)}\right] \nonumber \\
 =  2\text{Im}\text{Tr}\ln\left[i\mathcal{W}\text{sign}(G)\right],\label{eq:proof_matrix_pro}
\end{equation}
with
\begin{equation}
\text{sign}(G)\equiv \frac{\tanh({\frac{G}{2}})}{\sqrt{\tanh^2(\frac{G}{2})}},
\end{equation}
for gapped $G$. For gapless $G$, we notice that the amplitude of $\phi_W$ vanishes (see Eq.~\eqref{eq:S2}), so without loss of generality, we extend $\text{sign}(G)$ to encompass the gapless scenario by defining $\text{sign}(0)=0$ in the eigenbasis of $G$.
 Here, $i\mathcal{W}\,\text{sign}(G)$ is a unitary matrix,
and $\sqrt{\tanh^{2}\left(\frac{G}{2}\right)}$ an positive definite
matrix. The proof of Eq. (\ref{eq:proof_matrix_pro})
is delivered below. Hence, by taking the branch cut of the logarithm function on $[-\infty, 0]$,  we find 
\begin{equation}
\frac{\Delta\phi_{W}}{2\pi}=\frac{1}{\pi}\text{Im}\text{Tr}\ln\left[i\mathcal{W}\,\text{sign}(G)\right]=\frac{1}{2}\sum_{n}\text{sign}\left(q_{n}\right),\label{supp_eq:spectral_asym}
\end{equation}
with $q_{n}$ eigenvalue for $Q_W$. The second equality is based on the observation that the following formula
\begin{equation}
\text{Im}\int_{0}^{2\pi}\frac{dx}{2\pi}\partial_{x}\ln\left(\cos\frac{x}{2}+i\sin\frac{x}{2}\lambda\right)=\frac{1}{2}\text{sign}\left(\text{Re}\lambda\right),\ \text{for } \lambda \in \mathbb{C},
\end{equation}
singles out the real part of $\lambda$. Hence, together with Eq.~\eqref{eq:hermiticity} and Eq.~\eqref{eq:proof_matrix_pro}, we connect $\phi_W$ with the real part of the eigenvalues of $\mathcal{W}\sign(G)$ (recall that this matrix is normal, i.e. unitarily diagonalizable). But the real part of these eigenvalues coincides with the eigenvalues of $Q_W\equiv \frac{1}{2}\{\mathcal{W},\ \sign(G)\}$, which holds since $\mathcal{W}\sign(G)$ and $\sign(G)\mathcal{W} = [\mathcal{W}\sign(G)]^\dag$ are diagonalized by the same unitary transformation. Using  the interpolation $\text{Im}\text{Tr}\ln\left[\cos(\frac{x}{2})+i\sin(\frac{x}{2}) \mathcal{W}\,\text{sign}(G)\right]$ between $0$ and $\frac{1}{\pi}\text{Im}\text{Tr}\ln\left[i\mathcal{W}\,\text{sign}(G)\right]$, Eq.~\eqref{supp_eq:spectral_asym} follows.


\textit{Proof of Eq. (\ref{eq:proof_matrix_pro})}. 
It is based on the defining properties of the matrix logarithm and the Baker-Campbell-Hausdorff formula. For notational simplicity, we rewrite it as 
\begin{equation}
\text{Im}\text{Tr}\ln\left(U\times A\right)\overset{?}{=}\text{Im}\text{Tr}\ln U,\label{supp_eq:trln}
\end{equation}
where $U$ an unitary matrix, and $A$ a positive definite matrix. By definition of the matrix logarithm, we have
\begin{equation}
U=e^{\ln U},\ A=e^{\ln A}.
\end{equation}
Due to the Baker-Campbell-Hausdorff formula $e^{B_{1}}e^{B_{2}}=e^{B_{1}+B_{2}+\left[B_{1},\ \left[B_{1},\ \dots\right]\right]}$,
we find 
\begin{equation}
\text{Tr}\ln\left(U\times A\right)=\text{Tr}\ln e^{\ln U+\ln A+\left[\ln U,\ \left[\ln U,\ \dots\right]\right]}=\text{Tr}\left(\ln U+\ln A\right),
\end{equation}
and thus demonstrate Eq. (\ref{supp_eq:trln}). 


\section{Action Eq. (\ref{eq:effective_action}) from Dirac model \label{supp_sec:Dirac}}

We present  a detailed derivation of Eq. (\ref{eq:effective_action})
from a microscopic Dirac model in the presence of an external $U(1)$ gauge field, for the $\mathbb{Z}_{2}$ class (see \citep{huang2022prb} for the $\mathbb{Z}$ or $2\mathbb{Z}$ classes).
For concreteness, we present the derivation in $d=2$, the generalization
to higher dimension is straightforward. We start from the following continuum model, 
\begin{equation}
\hat{G}=\hat{G}_{0}+\int d^2\boldsymbol{x}[m\hat{\psi}^{\dagger}(\boldsymbol{x})\sigma^{z}\otimes\left(\boldsymbol{n}\cdot\boldsymbol{\tau}\right)\hat{\psi}(\boldsymbol{x})],\qquad 
\hat{G}_{0}=\int d^2 \boldsymbol{x}\hat{\psi}^{\dagger}(\boldsymbol{x})\left[\left(i\partial_{x}-A_{x}\right)\sigma^{x}\otimes\tau^{0}+\left(i\partial_{y}-A_{y}\right)\sigma^{y}\otimes\tau^{0}\right]\hat{\psi}(\boldsymbol{x}),
\end{equation}
where $\boldsymbol{n}$ is a unit vector, and $\boldsymbol{x}$ the 2-dimensional spatial coordinate. We will reserve the symbol $G_0$ ($G$) for the first-quantized  counterparts of $\hat{G}_0$ ($\hat{G}$), i.e., 
\begin{equation}
G_0 \equiv [i\partial_x - A_x(\boldsymbol{x})]\sigma^x \otimes \tau^0 +[i\partial_y - A_y(\boldsymbol{x})]\sigma^y \otimes \tau^0,
\end{equation}
containing two decoupled massless Dirac operators under $U(1)$ gauge field $(A_x, A_y)$,
and
\begin{equation}
G = G_0 +m\sigma^z \otimes (\boldsymbol{n}\cdot\boldsymbol{\tau}),
\end{equation}
involving a mass term $m \sigma^z \otimes \boldsymbol{n}\cdot\boldsymbol{\tau}$. As for the probe operator, we take $\mathcal{W}=\boldsymbol{s}\cdot\boldsymbol{\tau}\otimes \sigma^0$
with $\boldsymbol{s}$ a unit vector, such that
\begin{eqnarray}
\phi_{W} & \equiv & \text{arg}\text{Tr}\left(e^{-i\int d^2 \boldsymbol{x}\hat{\psi}^\dagger w\boldsymbol{s}\cdot\boldsymbol{\tau}\otimes \sigma^0 \hat\psi}e^{-\hat G}\right)\nonumber = \text{Im}\left\{ \text{Tr}\ln e^{-i\frac{1}{2}w\boldsymbol{s}\cdot\boldsymbol{\tau}\otimes \sigma^0}+\text{Tr}\ln\left[\cos\left(\frac{w}{2}\right)+i\sin\left(\frac{w}{2}\right)\boldsymbol{s}\cdot\boldsymbol{\tau}\otimes \sigma^0\tanh\left(\frac{G}{2}\right)\right]\right\},
\end{eqnarray}
where $w$ is homogenous, and $\text{Tr}$ is for tracing over both the internal and spatial space, while $\text{tr}$ is preserved for internal space. 

In the rest of this section, we will show that 
\begin{eqnarray}
\phi_{W}&=&\text{ch}_W \times\int d^{2}\boldsymbol{x}\mathcal{I}_{W}\left[w\right] \times \mathcal{N} , \qquad\mathcal{N}=\frac{1}{2\pi}\epsilon^{ij}\partial_i A_j, \nonumber\\
 \text{ch}_W &=& 2\,\frac{ \text{sign}(m)}{2}, \quad \mathcal{I}_{W}\left[w\right]=\text{Re}\left\{ -i\text{tr}\ln\left[\cos\left(\frac{w}{2}\right)+i\sin\left(\frac{w}{2}\right)\boldsymbol{s}\cdot\boldsymbol{\tau}\tanh\frac{ |m|\left(\boldsymbol{n}\cdot\boldsymbol{\tau}\right)}{2}\right]\right\}. \label{supp_eq:Dirac_IW}
\end{eqnarray}
Here, $\mathcal{N}$ is the topological index  density of $G_0$, which in the presence of the external magnetic field equals the specified expression. It
counts the number of zero modes associated
with $G_{0}$.
It is interesting to note that all the topological information is contained in $G_0$, while both the mass term (associated to $\boldsymbol{n}$) and the probe operator (associated to $\boldsymbol{s}$) appear exclusively in the function $\mathcal{I}_W$. The relation between the zero modes of the Dirac operator and the topological index is the content of the Atiyah-Singer index theorem; it holds in higher even space dimensions too, so that analogous results can be derived straightforwardly, 
inferred from the index theorem. 

To proceed, we will represent $\phi_{W}$ in the eigenbasis of $G_{0}$ (including inhomogenous external fields), aiming to demonstrate that $\phi_{W}$ is only from zero
modes of $G_{0}$. We first notice that $G_{0}$ has the following symmetries, 
\begin{equation}
\left\{ G_{0},\ \sigma^{z}\right\} =0,\qquad
\left[G_{0},\ \left(\boldsymbol{s}\cdot\boldsymbol{\tau}\otimes \sigma^0 \right)\right]=0,
\end{equation}
from which we infer that: \\
\begin{itemize}
\item Eigenstates of $G_{0}$ with \textit{non-zero}
eigenvalues appear in opposite eigenvalues pairs, namely $|\psi_{n}\rangle,\ \sigma^{z}|\psi_{n}\rangle$, with eigenvalues $\pm\lambda_n$. --- For later notational simplicity, we preserve the $n=0$ index for the zero-eigenvalue sector.\\
\item We can diagonalize $G_{0}$ and $\boldsymbol{s}\cdot\boldsymbol{\tau}\otimes \sigma^0$
simultaneously, and thus label $|\psi_{n}\rangle$ by $|\psi_{n}^{\pm}\rangle$
for the positive (negative) eigenvalue of $\boldsymbol{s}\cdot\boldsymbol{\tau}$.
\end{itemize}
Together, the contribution to $\phi_{W}$ from the eigenstates $|\psi_{n}^{\pm}\rangle$'s ($n\neq 0$)
is
\begin{equation}
\text{arg}\ \text{Tr}_n[ \cos\left(\frac{w}{2}\right)+i\sin\left(\frac{w}{2}\right)\boldsymbol{s}\cdot\boldsymbol{\tau}_n\tanh(\frac{1}{2}G_n) ],\label{supp:spectral_asym}
\end{equation}
where the subscript $n$ is to emphasize the restriction to the subspace spanned by $\{|\psi_n^\pm\rangle,\ \sigma^z |\psi_n^\pm\rangle\}$'s, e.g.,  $\text{Tr}_n$ for tracing over this subspace, and $\boldsymbol{s}\cdot\boldsymbol{\tau}_n$ and $G_n$ are 
\begin{equation}
\boldsymbol{s}\cdot\boldsymbol{\tau}_n\equiv \left(\begin{array}{cc}
\left(\begin{array}{cc}
1\\
 & -1
\end{array}\right) & 0\\
0 & \left(\begin{array}{cc}
1\\
 & -1
\end{array}\right)
\end{array}\right),
\end{equation}
and 
\begin{equation}
G_n \equiv \left(\begin{array}{cc}
\lambda_{n}\mathbb{1} & m\boldsymbol{n}\cdot\boldsymbol{\tau}\\
m\boldsymbol{n}\cdot\boldsymbol{\tau} & -\lambda_{n}\mathbb{1}
\end{array}\right).
\end{equation}

After these preparations, we are ready to demonstrate that $\phi_W$ receives no contributions from $n\neq 0$ states, i.e., Eq.~\eqref{supp:spectral_asym} vanishes for $n\neq0$. This is based on the following two properties: 1. $(\boldsymbol{s}\cdot \boldsymbol{\tau}_n)^2 = \mathbb{I}$; 2. $G_n$ possesses chiral symmetry, i.e., $\{G_n, M\}=0$ where $M\equiv \left(\begin{array}{cc}
0 & -i\mathbb{I}\\
i\mathbb{I} &0
\end{array}\right)$. In turn, we find $\text{Tr}_n[\cos(\frac{w}{2})+i\sin(\frac{w}{2}) \boldsymbol{s}\cdot \boldsymbol{\tau}_n \tanh(\frac{1}{2}G_n)]$ is real, i.e.,
\begin{eqnarray}
 &  & \text{Tr}_{n}\ln\left[\cos\left(\frac{w}{2}\right)+i\sin\left(\frac{w}{2}\right)\boldsymbol{s}\cdot\boldsymbol{\tau_n}\tanh\left(\frac{ G_n}{2}\right)\right]\nonumber \\
 & = & \text{Tr}_{n}[M (\boldsymbol{s}\cdot\boldsymbol{\tau}_n)]\ln \left[ \cos\left(\frac{w}{2}\right)+i\sin\left(\frac{w}{2}\right)\tanh\left(\frac{1}{2}G_{n}\right){\boldsymbol{s}\cdot\boldsymbol{\tau}}_{n}\right][M (\boldsymbol{s}\cdot\boldsymbol{\tau}_n)]^\dagger \nonumber \\
 & = & \text{Tr}_{n}\ln\left[\cos\left(\frac{w}{2}\right)+i\sin\left(\frac{w}{2}\right)\boldsymbol{s}\cdot\boldsymbol{\tau}_n\tanh\left(\frac{ G_n}{2}\right)\right]^\dagger.
\end{eqnarray}

Built upon results above, we conclude that $\phi_W$ is from the $n=0$ state (zero mode of $G_0$). To finally obtain Eq.~\eqref{supp_eq:Dirac_IW}, we proceed by utilizing the chiral symmetry of $G_0$, from which we can take the zero-mode to be eigenstates of $\sigma^z$, labeled by $\alpha=\pm $. In turn, $\phi_W$ becomes
\begin{equation}
\phi_W =\sum_{\alpha =\pm } \text{Re}(-i)\text{tr}\ln \left[\cos(\frac{w}{2})+  i\sin(\frac{w}{2}) \boldsymbol{s}\cdot\boldsymbol{\tau} \tanh(\frac{\alpha m \boldsymbol{n}\cdot \boldsymbol{\tau}}{2})\right],
\end{equation}
where we have used $[G_0, \boldsymbol{\tau}\otimes \sigma^0]=0$, and here, "$\text{tr}$" is over the two-dimensional $\boldsymbol{\tau}$ matrix. Together with the index theorem, we reproduce Eq.~\eqref{supp_eq:Dirac_IW}, i.e., 
\begin{eqnarray}
\phi_{W} & = & \text{ch}_W\times \left(n_{\alpha =+}-n_{\alpha = -}\right)\times \mathcal{I}_{W}\left[w\right]= \text{ch}_W\times\mathcal{I}_{W}\left[w\right]\times\left(\int\frac{d^{2}\boldsymbol{x}}{2\pi}\epsilon^{ij}\partial_{i}A_{j}\right),\ \ \text{ch}_W = 2\,\frac{ \text{sign}(m)}{2},
\end{eqnarray}
and 
\begin{equation}
n_{\alpha = +}-n_{\alpha = -}\equiv \int d^{2}\boldsymbol{x} \mathcal{N}\in \mathbb{Z}.
\end{equation}
In particular, we notice that $\mathcal{I}_W$ satisfies
 \begin{equation}
\mathcal{I}_W|_{w}^{w+2\pi}=\pm 2 \pi,
\end{equation}
where the $\pm$ originates from the sign of $\text{tr}[(\boldsymbol{s}\cdot \boldsymbol{\tau})\cdot(\boldsymbol{n}\cdot \boldsymbol{\tau})]$. 
$\text{ch}_W$ is the spin Chern number \cite{sheng2006prl, roy2009prbz2, prodan2009prb} for our massive Dirac model: This can be seen clearly in the limit of $\boldsymbol{n}\cdot \boldsymbol{\tau} =\tau^z$, where $G$ consists of two Dirac models with opposite mass, and thus opposite Chern number, i.e., $\pm \frac{1}{2}\text{sign}(m)$ \cite{ryu2010njp}. The half-integer valuedness is attributed to the Dirac model, which can be cured by proper regularization \cite{redlich1984prl, ryu2010njp,huang2022prb}.  These together produce  $\text{ch}_W = \frac{1}{2}\sign(m) - [-\frac{1}{2}\sign(m) ] =2\frac{\text{sign}(m)}{2}$.

Finally, we highlight that our action is readily generalized to weakly inhomogenous $w$, via gradient expansion as derived in \cite{huang2022prb}, by replacing constant $w$ with $w[\boldsymbol{x}]$.

\begin{table}
\begin{tabular}{|c|c|c|c|c|c|c|c|c|c|c|c|}
\hline 
Symmetry class & \multicolumn{1}{c}{Symmetry} &  & $d=0$ & $1$ & $2$ & $3$ & 4 & 5 & 6 & 7 & 8\tabularnewline
\hline 
\hline 
AI & $\mathcal{T}^{2}=1$ &  & $\mathbb{Z}$ & $0$ & $0$ & $0$ & ${\color{brown}2\mathbb{Z}}$ & $0$ & ${\color{blue}\mathbb{Z}_{2}}$ & ${\color{purple}\mathbb{Z}_{2}}$ & $\mathbb{Z}$\tabularnewline
\hline 
BDI & $\mathcal{T}^{2}=1$ & $\mathcal{C}^{2}=1$ & ${\color{purple}\mathbb{Z}_{2}}$ & $\mathbb{Z}$ & $0$ & $0$ & 0 & ${\color{brown}2\mathbb{Z}}$ & $0$ & ${\color{blue}\mathbb{Z}_{2}}$ & ${\color{purple}\mathbb{Z}_{2}}$\tabularnewline
\hline 
D &  & $\mathcal{C}^{2}=1$ & ${\color{blue}\mathbb{Z}_{2}}$ & ${\color{purple}\mathbb{Z}_{2}}$ & $\mathbb{Z}$ & $0$ & $0$ & 0 & ${\color{brown}2\mathbb{Z}}$ & $0$ & ${\color{blue}\mathbb{Z}_{2}}$\tabularnewline
\hline 
DIII & $\mathcal{T}^{2}=-1$ & $\mathcal{C}^{2}=1$ & $0$ & ${\color{blue}\mathbb{Z}_{2}}$ & ${\color{purple}\mathbb{Z}_{2}}$ & $\mathbb{Z}$ & $0$ & $0$ & 0 & ${\color{brown}2\mathbb{Z}}$ & $0$\tabularnewline
\hline 
AII & $\mathcal{T}^{2}=-1$ &  & ${\color{brown}2\mathbb{Z}}$ & $0$ & ${\color{blue}\mathbb{Z}_{2}}$ & ${\color{purple}\mathbb{Z}_{2}}$ & $\mathbb{Z}$ & $0$ & $0$ & 0 & ${\color{brown}2\mathbb{Z}}$\tabularnewline
\hline 
CII & $\mathcal{T}^{2}=-1$ & $\mathcal{C}^{2}=-1$ & $0$ & ${\color{brown}2\mathbb{Z}}$ & $0$ & ${\color{blue}\mathbb{Z}_{2}}$ & ${\color{purple}\mathbb{Z}_{2}}$ & $\mathbb{Z}$ & $0$ & $0$ & 0\tabularnewline
\hline 
C &  & $\mathcal{C}^{2}=-1$ & $0$ & $0$ & ${\color{brown}2\mathbb{Z}}$ & $0$ & ${\color{blue}\mathbb{Z}_{2}}$ & ${\color{purple}\mathbb{Z}_{2}}$ & $\mathbb{Z}$ & $0$ & $0$\tabularnewline
\hline 
CI & $\mathcal{T}^{2}=1$ & $\mathcal{C}^{2}=-1$ & $0$ & $0$ & $0$ & ${\color{brown}2\mathbb{Z}}$ & 0 & ${\color{blue}\mathbb{Z}_{2}}$ & ${\color{purple}\mathbb{Z}_{2}}$ & $\mathbb{Z}$ & $0$\tabularnewline
\hline 
\end{tabular}

\caption{Symmetry shifting pattern induced by probe operator: 1. In even spatial dimension: ${\color{blue}\mathbb{Z}_2}$ and ${\color{purple}\mathbb{Z}_2}$ to ${\color{brown}2\mathbb{Z}}$; 2. Dimensional reduction, from even spatial dimension to odd spatial dimension: ${\color{blue}\mathbb{Z}_2}$ (even D) to ${\color{blue}\mathbb{Z}_2}$ (odd D);  $\mathbb{Z}$ (even D) to ${\color{purple}\mathbb{Z}_2}$ (odd D).
\label{supp_tab:10_class}}
\end{table}

\section{$\mathbb{Z}_2$ invariant from the freedom of choosing $\mathcal{W}$ }

Here we shall focus on the $\mathbb{Z}_{2}$ class in even spatial dimensions, and elucidate the emergence of its $\mathbb{Z}_{2}$ invariant through variation of $\mathcal{W}$. To this end, we shall first demonstrate that opting for the naive choice $\mathcal{W}=\mathbb{I}$ results in a vanishing $\frac{\Delta\phi_{W}}{2\pi}$. This observation prompts us to seek for an alternative $\mathcal{W}$. We construct it such that the associated $Q_{W}$ falls in a $2\mathbb{Z}$ class. It turns out that the choice of $\mathcal{W}$ is not unique, which leads to a reduction form $2\mathbb{Z}$ to $\mathbb{Z}_{2}$ invariant. For concreteness, we shall focus on external $U\left(1\right)$ gauge fields in this section. 

\subsection{Vanishing $\Delta\phi_{W}$ for $\mathcal{W}=\mathbb{I}$ \label{supp_sec:vanishing_spec}}

We first show that symmetry enforces vanishing $\frac{\Delta\phi_{W}}{2\pi}|_{\mathcal{W}=\mathbb{I}}=0$ in the $\mathbb{Z}_{2}$ class in even spatial dimensions. We observe that:
\begin{itemize}
\item The periodic table exhibits the following symmetry pattern for the $\mathbb{Z}_{2}$ classes: In $d = 4k$ ($d=4k+2$) spatial dimensions $(k\in\mathbb{Z})$, all the $\mathbb{Z}_{2}$ classes possess particle-hole (time-reversal) symmetry, i.e., $\mathcal{S}G\mathcal{S}^{-1}=(-1)^{d/2 + 1} G$, with $\mathcal{S}$ for $\mathcal{C}$ (or $\mathcal{T}$).

\item Under an external field with strength tensor $F_{\mu\nu}\equiv \partial_\mu A_\nu -\partial_\nu A_\mu$, the spectral asymmetry formula in Eq.~\eqref{eq:spec_asym} in the main text becomes, 
\begin{equation}
\frac{\Delta\phi_W}{2\pi}=\text{ch}_W \int \mathfrak{C},\ \text{and}\ \mathfrak{C}=\frac{\epsilon^{0i_1 i_2\dots i_{d-1}i_{d}}}{(2\pi)^{d/2}(d/2)!}\partial_{i_1}A_{i_2}\dots \partial_{i_{d-1}}A_ {i_{d}},
\label{supp_eq:spec_asym_magnetic_field}
\end{equation}
where "$\int$" denotes the integration over the spatial coordinates. Expressed in terms of a magnetic field, we have $\mathfrak{C} =(\frac{B}{2\pi})^{d/2}$.
\end{itemize}
Consequently, symmetry necessitates vanishing spectral asymmetry: On the one hand, 
\begin{equation}
\frac{\Delta \phi_W}{2\pi} = \frac{1}{2}\text{Tr}[\sign (G[A])] = \text{ch}_W \int \mathfrak{C}, \label{supp_eq:delta_phi1}
\end{equation}
which results from Eq.~\eqref{supp_eq:spectral_asym}, by taking $\mathcal{W}=\mathbb{I}$. On the other hand, from time-reversal (particle-hole) symmetry with odd (even) $d/2$, 
\begin{equation}
\frac{\Delta \phi_W}{2\pi} = (-1)^{d/2+1} \frac{1}{2}\text{Tr}[\mathcal{S}\sign (G[-A])\mathcal{S}^{-1}] = - \text{ch}_W \int \mathfrak{C}, \ \text{with}\ \mathcal{S} = \mathcal{C} \ \text{or} \ \mathcal{T},\label{supp_eq:delta_phi2}
\end{equation}
where for odd $d/2$, the minus sign in the second equality is from $A\rightarrow -A$ in $\mathfrak{C}$; for even $d/2$, it is from particle-hole symmetry [$\mathcal{C}G\mathcal{C}^{-1}= (-1)G$], encoded in the prefactor $(-1)^{d/2+1}$. By comparing Eq.~\eqref{supp_eq:delta_phi1} with Eq.~\eqref{supp_eq:delta_phi2}, we conclude that 
\begin{equation}
\frac{\Delta \phi_W}{2\pi}=0.
\end{equation}

Alternatively, the vanishing spectral asymmetry can be attributed to \textit{symmetric pairs}, which is a crucial concept in the following. By a symmetric pair, we refer to two eigenvalues of $\sign(G)$ with opposite sign, whose associated eigenstates are related to each other by a symmetry transformation $\mathcal{S}$. Indeed, imposing of a symmetry compels a pairwise division within the spectrum of 
$G$: The symmetry transformation
\begin{equation}
\mathcal{S}G[A]\mathcal{S}^{-1}= (-1)^{d/2+1} G[-A]
\end{equation}
implies that for every eigenstate $|\psi[A]\rangle$ with eigenvalue $g[A]$, its symmetric counterpart $\mathcal{S}^{-1}|\psi[-A]\rangle$ has eigenvalue $(-1)^{(d/2+1)}g[-A]$. Thus, the spectrum comprises pairs $\{g[A],\ (-1)^{d/2+1}g[-A]\}$, each with a level degeneracy $|\int \mathfrak{C}|$. Furthermore, motivated by the spectral asymmetry formula  Eq.~\eqref{supp_eq:spec_asym_magnetic_field}, 
we can restrict our attention to states  contributing terms linear in $\int \mathfrak{C}$, whose eigenvalues must satisfy
\begin{equation}
\text{sign}(g[A])=\text{sign}(\mathfrak{C})\implies  \text{sign}(g[A]) |\int \mathfrak{C}|=\int \mathfrak{C},
\end{equation}
where $|\int \mathfrak{C}|$ is from level degeneracy, and where we have neglected a possible $A$-field independent sign in $g[A]$, as it is irrelevant for our results. Consequently, eigenvalues from these symmetric pairs must exhibit opposite signs, i.e., 
\begin{equation}
\text{sign}\{g[A]\times (-1)^{d/2+1}g[-A]\}=\sign\left[(-1)^{d+1}\right]<0,
\end{equation}
from which we conclude that the external field activated spectral asymmetry vanishes, due to the presence of symmetric pairs with opposite sign eigenvalues. 

So far, we have shown that symmetry enforces symmetric pairs, and thus a vanishing spectral asymmetry (for $\mathcal{W}=\mathbb{I}$), due to a cancellation effect. To cure this problem, i.e. to obtain a finite signal quantitatively probing the underlying topology, our strategy is to reverse the relative sign for these symmetric pairs, implemented via introducing a $\mathcal{W}$ matrix such that $Q_W\equiv \frac{1}{2}\{\sign(G),\ \mathcal{W}\}$ contains non-zero spectral asymmetry, and belongs to the $2\mathbb{Z}$ class.

\subsection{Symmetry constraints for $\mathcal{W}$ \label{supp_sec:sym_pairs}}
To place $Q_W$
in the $2\mathbb{Z}$ class in the same dimension (brown in Tab.~\ref{supp_tab:10_class}), possessing an even integer valued spectral asymmetry (with $\mathcal{W}=\mathbb{I}$), we equip $\mathcal{W}$
with a symmetry,
\begin{enumerate}
\item $G$ without chiral symmetry (blue in Tab.~\ref{supp_tab:10_class}): For $G$ with $\mathcal{T}$ (or $\mathcal{C}$) symmetry, $\mathcal{T}\mathcal{W}\mathcal{T}^{-1}=-\mathcal{W}$ {(}or
$\mathcal{C}\mathcal{W}\mathcal{C}^{-1}=-\mathcal{W}${)}.\label{supp_item:nchiral}
\end{enumerate}
To see this, we note that the symmetries of $Q_W$ are
\begin{equation}
\begin{cases}
\mathcal{T}G\mathcal{T}^{-1}=G\\
\mathcal{T}=U_{T}\mathcal{K},\ \mathcal{T}^{2}=\pm1
\end{cases}\rightarrow\begin{cases}
\mathcal{C}_W Q_W \mathcal{C}_W^{-1}=-Q_W\\
\mathcal{C}_W=U_{T}\mathcal{K},\ \mathcal{C}_W^{2}=\pm1
\end{cases},
\end{equation}
and 
\begin{equation}
\begin{cases}
\mathcal{C}G\mathcal{C}^{-1}=-G\\
\mathcal{C}=U_{C}\mathcal{K},\ \mathcal{C}^{2}=\pm1
\end{cases}\rightarrow\begin{cases}
\mathcal{T}_W Q_W \mathcal{T}_W^{-1}=Q_W\\
\mathcal{T}_W=\text{sign}(G)U_{C}\mathcal{K},\ \mathcal{T}_W^{2}=\mp1
\end{cases}.
\end{equation}
which confirms that $Q_W$ belongs to the $2\mathbb{Z}$ class.
The property $\mathcal{T}_W^{2}=\mp1$ in the second equation results from a special symmetry: 
\begin{equation}
\left[Q_W,\ \sign(G)\right]=0 \implies \mathcal{T}^{2}_W=\sign(G) U_{C}\mathcal{K}\sign(G)U_{C}\mathcal{K}=-\left(U_{C}\mathcal{K}\right)^{2}.
\end{equation}
and, due to this
symmetry, the inherited $\mathcal{T}_W$ symmetry can satisfy either $\mathcal{T}^{2}_W=+1,\ -1$ by dressing up with $\sign(G)$,  i.e., $\sign{G}U_C\mathcal{K}$, and our choice here is to ensure $Q_W$ in the $2\mathbb{Z}$ class. Interestingly, this exhibits a shifting pattern, i.e., $\mathcal{T}(\mathcal{C})\rightarrow \mathcal{C}_W (\mathcal{T}_W)$, which will be utilized to construct the $\mathcal{W}$ matrix for the chiral symmetry $\mathbb{Z}_2$ class, discussed below. 
\begin{enumerate}[label=2.]
\item $G$ with chiral symmetry (purple  in Tab.~\ref{supp_tab:10_class}):\label{supp_item:chiral} \\
(a) $\mathcal{W}$ anticommutes with the
chiral symmetry generator; \\
(b) $\mathcal{T}\mathcal{W}\mathcal{T}^{-1}=-\mathcal{W}$
(or $\mathcal{C}\mathcal{W}\mathcal{C}^{-1}=-\mathcal{W}$) in $4k+2$
(or $4k$) spatial dimensions, with $k\in\mathbb{Z}$.
\end{enumerate}
We focus on the chiral symmetric $\mathbb{Z}_2$ class in even spatial dimensions (blue in Tab.~\ref{supp_tab:10_class}). To this end, we observe that they possess both particle-hole and time-reversal symmetry, while the  $2\mathbb{Z}$ class in the same dimension (brown in Tab.~\ref{supp_tab:10_class}) exhibits either particle-hole or time-reversal symmetry. To bridge this discrepancy, we require $\mathcal{W}$ to anticommute with the chiral symmetry generator ((a) in \ref{supp_item:chiral}), such that the resulting $Q_W$ is not chiral symmetric. Adding to this, we require $Q_W$ to share the same symmetry as the $2\mathbb{Z}$ class, from which one can straightforwardly infer (b) in \ref{supp_item:chiral}, via the symmetry shifting pattern (i.e., $\mathcal{T}(\mathcal{C})\rightarrow \mathcal{C}_W (\mathcal{T}_W)$).

Finally, we outline the construction of $\mathcal{W}$ matrix in odd spatial dimensions, but postpone the details to the later Sec. \ref{supp_sec:dr}, where a parallel scenario unfolds for the $\mathbb{Z}_2$ classes: 
\begin{itemize}
    \item For the non-chiral symmetric $\mathbb{Z}_2$ class (purple $\mathbb{Z}_2$ in Tab.~\ref{supp_tab:10_class}),  $\mathcal{W}=\mathbb{I}$ as $\phi_W$ can be non-zero.
    \item For the chiral symmetric class (blue $\mathbb{Z}_2$ in Tab.~\ref{supp_tab:10_class}), symmetry dictates vanishing $\phi_W$, necessitating an alternative choice of $\mathcal{W}$ matrix. This is achieved via the method of dimensional reduction, resulting in a $\mathcal{W}$ similar to its even spatial dimensional parent (i.e., blue $\mathbb{Z}_2$ in even spatial dimensions in Tab.~\ref{supp_tab:10_class}). 
\end{itemize}

\subsection{$\mathbb{Z}_{2}$ invariant from the freedom in choosing $\mathcal{W}$ \label{supp_sec:symmetry_pairs}}

The construction of $\mathcal{W}$ above introduces ambiguities in the selection of 
$\mathcal{W}$, to be addressed now. The ambiguity is noticed based on the observation that smooth deformations of $\mathcal{W}$ or $G$ lead to \textit{a transfer of symmetric pairs of $\sign(G)$} between the positive/negative eigenvalue sectors of $Q_W$ (denoted as $Q_W^{(\pm)}$), such that a symmetric pair of eigenstates in the $Q_W^{(\pm)}$ subspace is relocated to $Q_W^{(\mp)}$: Here, we have used the property that $Q_W$ commutes with $\sign(G)$, so eigenstates associated with a symmetric pair of $\sign(G)$ are also eigenstates of $Q_W$, belonging to the same positive/negative eigenvalue sector of $Q_W$ (see Sec.~\ref{supp_sec:derivations_sym_pairs_1+} for derivations). 
From this, we demonstrate that: 
\begin{enumerate}
\item The ensuing change of spectral asymmetry counts twice the number of transferred symmetric pairs. \label{supp_item1:S3c}
\item The condition that $G$ be gapped necessitates an even number of transferred symmetric pairs.\label{supp_item2:S3c}
\item $\mathcal{W}$ and $G$ are independent, so a smooth deformation of $\mathcal{W}$ changes the spectral asymmetry by $4\mathbb{Z}$, from which we establish a $\mathbb{Z}_{2}= 2\mathbb{Z} \mod 4$ invariant. \label{supp_item3:S3c}
\end{enumerate}

To this end, we first collect properties of $Q_W$, and introduce notation for later convenience:
\begin{itemize}
\item $\left[\text{sign}\left(G\right),\ Q_{W}\right]=0$, so we can diagonalize these two matrices simultaneously, i.e.,
\begin{equation}
Q_{W}=Q_{W}^{\left(1,\ +\right)}\oplus Q_{W}^{\left(1,\ -\right)}\oplus Q_{W}^{\left(2,\ +\right)}\oplus Q_{W}^{\left(2,\ -\right)}, 
\end{equation}
where $\pm$ (or $1,\ 2$) for the positive/negative sign of eigenvalues associated with $Q_W$ (or $\text{sign}(G)$), and $Q_W^{(1/ 2,\ \pm)}$ for the subspace consisting of corresponding eigenvectors.
\item For notational simplicity, we introduce $\lambda_{n}^{\left(1/2,\ \pm\right)}=1$
to enumerate states in the $Q_{W}^{\left(1/2,\ \pm\right)}$ subspace. 
\end{itemize}
Leveraging the properties listed above and employing Eq.~\eqref{supp_eq:spectral_asym}, the spectral asymmetry associated with $Q_W$ is 
\begin{equation}
\frac{\Delta\phi_W}{2\pi} = \frac{1}{2}\sum_n \{[\lambda_n^{(1, +)}+\lambda_n^{(2, +)}] -[\lambda_n^{(1, -)}+\lambda_n^{(2, -)}] \},\label{supp_eq:spec_asym_W}
\end{equation}
Here, $\lambda_n^{(1,\ +)}$ and  $\lambda_n^{(2,\ +)}$ (as well as $\lambda_n^{(1,\ -)}$ and  $\lambda_n^{(2,\ -)}$) form a symmetric pair (see Sec.~\ref{supp_sec:derivations_sym_pairs_1+} for derivation), which ensures that the spectral asymmetry associated with $G$ vanishes, i.e., 
\begin{equation}
 \sum_n \{[\lambda_n^{(1, +)}-\lambda_n^{(2, +)}] +[\lambda_n^{(1, -)}-\lambda_n^{(2, -)}] \} =0.
\end{equation}

Now we are ready to establish the $\mathbb{Z}_2$ invariant. To this end, we show that under the  constraint that $G$ remain gapped,  $\frac{\Delta\phi_W}{2\pi}$ undergoes a change of $4\mathbb{Z}$ under a smooth deformation of $\mathcal{W}$ or $G$. 
To achieve this, we first note that the change in spectral asymmetry is equal to twice the number of transferred symmetric pairs, i.e., \ref{supp_item1:S3c} listed above (see Tab.~\ref{supp_Tab:4Z_gaplessG}): We assign the number of states in $Q_W^{(1,\ +)}$ 
and $Q_W^{(1,\ -)}$ as $r\in \mathbb{Z}$ and $s-r \in \mathbb{Z}$, respectively, whose ensuing spectral asymmetry is $\frac{\Delta\phi_W}{2\pi}=2r-s$. After transferring $t$ symmetric pairs between the $Q_W^{(\pm)}$ sectors, i.e., $r\rightarrow r+t$ and $s-r\rightarrow s-r-t$, the spectral asymmetry changes by $2t\in 2\mathbb{Z}$. 

Furthermore, we observe that transferring an odd number of symmetric pairs necessitates a gapless $G$ (i.e., \ref{supp_item2:S3c} listed above). Illustratively, consider a $\{\lambda_n^{(1/2,\ +)}\}$ sub-block of $G$ (denoted by $G_{(s)}$), featuring a single symmetric pair, i.e., $G_{(s)}= \xi \tau^z$, where $\tau^z$ signifies the symmetric pair with opposite eigenvalues in $\sign(G)$. Moreover, $G_{(s)}$ exhibits "particle-hole" symmetry, i.e., $\mathcal{S}_{(s)}G_{(s)}[A]\mathcal{S}_{(s)}^{-1}=-G_{(s)}[A]$ and $\mathcal{S}_{(s)}=\mathcal{K}\tau^x$ (see Sec. \ref{supp_sec:example} for a concrete example). This reflects the underlying time-reversal/particle-hole symmetry, enforcing symmetric pairs of $\sign(G)$ as elaborated in Sec.~\ref{supp_sec:vanishing_spec}. 
After these preparations, let us implement a smooth deformation. 
Symmetry $\mathcal{S}_{(s)}$ dictates $\mathcal{W}_{(s)}$ to be $\mathcal{W}_{(s)}=\pm \tau^z$, satisfying $\mathcal{S}_{(s)}\tau^z\mathcal{S}_{(s)}^{-1}=-\tau^z$. Consequently, $Q_{W,\ (s)}=\sign(\xi)\tau^0$ when $\mathcal{W}_{(s)}=\tau^z$, where $\sign(\xi)$ identifies the belonging to the sectors $Q_W^{(\sign(\xi))}$.
Hence, the transfer of a symmetric pair necessitates tuning $\xi$ from positive to negative, closing the gap in $G$ at $\xi=0$ and inducing a spectral asymmetry change of $2$. This underscores the requisite gaplessness in $G$ for symmetric pair transfer.

In contrast, gapped $G$ mandates an even number of transferred pairs (i.e., \ref{supp_item3:S3c} listed above). Consider an illustrative case of a doublet: $G_{(s)}=\xi \tau^z \otimes \sigma^0$ and $\mathcal{S}_{(s)}=\mathcal{K}\tau^x \otimes \sigma^0$. We transfer this symmetric pair doublet between $Q_W^{\pm}$ through smooth deformations of either $G_{(s)}$ or $\mathcal{W}_{(s)}$, while keeping $G$ gapped. For the former scenario, we deform $G_{(s)}$ by adding an extra mass term allowed by symmetry, i.e., $\tau^y \otimes \sigma^y$, rendering $G_{(s)}=\xi(\cos\theta \tau^z\otimes \sigma^0 + \sin\theta \tau^y\otimes \sigma^y)$, with $\theta$ parameterizing the deformation. Taking $\mathcal{W}_{(s)} =\tau^z \otimes \sigma^0$, we find that the asymmetry matrix is $Q_{W,\ (s)}=\sign(\xi)\cos\theta \tau^0 \otimes \sigma^0$. Thus, a symmetric pair doublet can be transferred by tuning $\theta$, keeping $G$ gapped. Alternatively, the transfer can be achieved by deforming $\mathcal{W}$: Using $G_{(s)}=\xi \tau^z\otimes \sigma^0$ and $\mathcal{W}_{(s)} = \cos\theta\tau^z \otimes \sigma^0 + \sin\theta \tau^y\otimes \sigma^y$, we again infer that $Q_{W,\ (s)}=\sign(\xi)\cos\theta \tau^z \otimes \tau^0$. Hence, tuning $\theta$ enables the transfer of a symmetric pair doublet with $G$ remaining gapped, supporting our earlier assertion.

Together, we conclude that under the constraint of gapped $G$, the spectral asymmetry changes by $4\mathbb{Z}$, from transferring even number of symmetric pairs. Finally, it is amused to observe that the even/odd effect of symmetric pairs mirrors the $\mathbb{Z}_2$ classification in zero-dimensional D-class insulators. Specifically, an even number of D-class insulators (e.g., $G_{(s)}=\xi\tau^z$ representing one specific D insulator here), are adiabatically connected with spectral gap open, and thus belongs to the topologically trivial phase.


\begin{table}
\begin{tabular}{|c|c|c|c|c|c|c|}
\hline 
Gapless $G$ & $Q_{W}^{\left(1,\ +\right)}$ & $Q_{W}^{\left(2,\ +\right)}$ & $Q_{W}^{\left(1,\ -\right)}$ & $Q_{W}^{\left(2,\ -\right)}$ & $\frac{\Delta\phi_{W}}{2\pi}$ & Changes in $\frac{\Delta\phi_{W}}{2\pi}$\tabularnewline
\hline 
\hline 
$\#$ of states (before) & $r$ & $r$ & $s-r$ & $s-r$ & $2r-s$ & $2t\in2\mathbb{Z}$\tabularnewline
\cline{1-6} \cline{2-6} \cline{3-6} \cline{4-6} \cline{5-6} \cline{6-6} 
$\#$ of states (after) & $r+t$ & $r+t$ & $s-r-t$ & $s-r-t$ & $2r-s+2t$ & \tabularnewline
\hline 
\end{tabular}

\caption{Changes of $\frac{\Delta\phi_W}{2\pi}$ from transferring symmetric pairs.  
$Q_{W}$ contains $2s$ states.
\textit{Before} crossing zeros, the number of states in $Q_{W}^{\left(1,\ \pm\right)}$ (and $Q_{W}^{\left(2,\ \pm\right)}$) is $r\in\mathbb{Z}$. \textit{After} crossing, the number of states in $Q_W^{(1,\ +)}$ becomes $r + t\in\mathbb{Z}$. Consequently, the difference in the spectral asymmetry is $2t\in 2\mathbb{Z}$.\label{supp_Tab:4Z_gaplessG}}

\end{table}




\subsubsection{
Symmetric pairs of $Q_W^{\pm}$ \label{supp_sec:derivations_sym_pairs_1+}}
For $\sign(G)$, we have established the existence of eigenvalue pairs with \textit{opposite} signs, linked by symmetry.  Extending our investigation to $Q_W^{+}$ ($Q_W^{-}$), we observe a parallel phenomenon: eigenvalue pairs sharing the \textit{same} sign, interrelated through symmetry, which is referred to as \textit{symmetric pairs of $Q_W^{+}\ (Q_W^{-})$}. To show this, we  adopt a strategy akin to that presented in Sec.~\ref{supp_sec:vanishing_spec}, utilizing both the spectral asymmetry formula Eq.~\eqref{supp_eq:spec_asym_magnetic_field} and the symmetry relation $\mathcal{S} Q_W[A]\mathcal{S}^{-1} = (-1)^{d/2+2} Q_W[-A]$. This entails

\begin{enumerate}
\item The spectral asymmetry formula, expressed as $\frac{\Delta\phi_W}{2\pi}=\text{ch}_W \int \mathfrak{C}$, implies that it is sufficient to focus on eigenvalues of $Q_W$ (denoted by $q[A]$). These eigenvalues adhere to $\sign q[A] = \sign \mathfrak{C}$, up to a possible $A$-field-independent sign.
\item The symmetry relation $\mathcal{S} Q_W[A]\mathcal{S}^{-1} = (-1)^{d/2+2} Q_W[-A]$ dictates that for the eigenstate of $Q_W$ (denoted by $|\psi[A]\rangle$) with eigenvalue $q[A]$, its symmetry transformed counterpart (denoted by $\mathcal{S}^{-1}|\psi[-A]\rangle$) has eigenvalue $(-1)^{d/2+2} q[-A]$.
\end{enumerate}
Consequently, we conclude that the eigenvalues of $|\psi[A]\rangle$ and $\mathcal{S}^{-1}|\psi[-A]\rangle$ share the same sign, i.e., 
\begin{equation}
\sign\{q[A]\times (-1)^{d/2+2} q[-A]\} =(-1)^{d+2}>0,
\end{equation}
and thereby confirm the existence of symmetric pairs of $Q^{(+)}_W$ and of $Q^{(-)}_W$. We re-emphasize that, the symmetric pairs of $\sign{G}$ have the opposite sign, the symmetric pairs of $Q^{(\pm)}_W$ have the same sign.

\subsubsection{An example for the symmetric pairs of $\sign(G)$ and $Q_W$} \label{supp_sec:example}

Here, we shall illustrate symmetric pairs discussed above in a concrete example, from which we reveal an emergent particle-hole symmetry, and dimensional reduction to $0$-dimensional class-D insulators.  

We consider the two-dimensional time-reversal symmetric Dirac model in Sec.~\ref{supp_sec:Dirac}, and derive the Hamiltonian consisting of one symmetric pair of $\sign(G)$,
\begin{equation}
G = G_0 +m\sigma^z \otimes (\boldsymbol{n}\cdot \boldsymbol{\tau}),\ \boldsymbol{n}=(0,\ 0,\ 1),
\end{equation}
with 
\begin{equation}
G_0 \equiv [i\partial_x - A_x(\boldsymbol{x})]\sigma^x \otimes \tau^0 +[i\partial_y - A_y(\boldsymbol{x})]\sigma^y \otimes \tau^0,\ \text{and}\ \mathcal{T}=\mathcal{K} \sigma^y \otimes \tau^x,
\end{equation}
where the choice of $\boldsymbol{n}$ is made for simplicity. In Sec.~\ref{supp_sec:Dirac}, we have demonstrated that under magnetic fields, the spectral asymmetry is from zero modes of $G_0$, in particular the difference in chiral zero modes (i.e., chiral matrix $\sigma^z\otimes \tau^0$). Hence, it is sufficient to focus on this zero mode sector of $G_0$ (i.e., by setting $G_0$ to zero), rendering an effective Hamiltonian in the symmetric pair basis (with degeneracy $|\int \mathfrak{C}|$), 
\begin{equation}
G_{(s)}[A] =m\times \text{sign}(\mathfrak{C}) \tau^z,\  \text{with}\ \mathcal{S}_{(s)}G_{(s)}[A]\mathcal{S}_{(s)}^{-1} =-G_{(s)}[A],\ \text{and}\ \mathcal{S}_{(s)}=\mathcal{K}\tau^x,
\end{equation}
where the "chiral matrix" $\sigma^z$ drops out by restricting to one chiral sector, because we are only interested in the difference of chiral modes. Clearly, this effective model possesses one symmetric pair of $\sign(G)$, signaled by the emergent "particle-hole symmetry" $\mathcal{S}_{(s)}=\mathcal{K}\tau^x$ ($\mathcal{S}_{(s)}^2=1$). The latter is inherited from its parent $\mathcal{T}$ symmetry, such that $\sigma^{y}$ drops out: The effect of $\sigma^y$ (upon $G$) is to reverse both the sign of magnetic fields and mass, and thus acts trivially on $G_{(s)}$. Physically speaking, this is because magnetic field significantly constrains the "spin" degree of freedom, such that the operation upon "spin" (i.e., $\sigma^y$ in $\mathcal{T}$) is irrelevant for the effective model. Finally, we remark that the magnetic field renders a dimensional reduction from $2$ dimensional time-reversal massive Dirac fermions (i.e., AII insulator) to $0$ dimensional particle-hole massive ones (i.e., D insulator). 

Built upon this, we can further construct the ensuing asymmetry matrix $Q_W$, containing one symmetric pair. This is based on the observation that in the basis of symmetric pairs, the only $\mathcal{W}$ allowed by symmetry is,
\begin{equation}
\mathcal{W}_{(s)} = \tau^z,\ \text{with}\ \mathcal{S}_{(s)}\mathcal{W}_{(s)}\mathcal{S}_{(s)}^{-1} = -\mathcal{W}_{(s)}, 
\end{equation}
from which one can infer the effective $Q_W$, i.e.,  
\begin{equation}
Q_{W,\ (s)} = \text{sign}(m \mathfrak{C}) \tau^0,
\end{equation}
consistent with our discussion in Sec. \ref{supp_sec:sym_pairs}, with $\xi = m\times \text{sign}(\mathfrak{C})$.

\section{Results in odd spatial dimension from dimensional reduction \label{supp_sec:dr}}
Here, we generalize our even dimensional results to odd spatial dimensions, via the standard method of dimensional reduction \cite{qi2008prb, qi2011rmp, huang2022prb}. This is achieved in two steps: (i). formulating the even dimensional effective action for $\phi_W$ in terms of external fields, and (ii). deriving the ensuing odd spatial dimensional one, by integrating out one spatial dimensions. From this odd dimensional $\phi_W$, we can infer the ensuing order parameter, which falls into two categories, $\mathbb{Z}_2$, or $\mathbb{Z}$ ($2\mathbb{Z}$), determined by chiral symmetry.

We now implement this protocol, so as to derive $\phi_W$ in odd spatial dimensions. To this end, we start from an effective action for $\phi_W$ in $d=2n$ spatial manifold $\mathcal{M}^{(2n)}=\mathcal{M}^{(2n-1)}\times S^1$, i.e.,
\begin{equation}
\phi_W\equiv \text{ch}_W\times \int_{\mathcal{M}^{(2n)}} d^{2n}\boldsymbol{x} \mathcal{I}_W[w(\boldsymbol{x})] \times \mathfrak{C}^{(2n)},\label{supp_eq:effective_action}
\end{equation}
where the superscript $(2n)$ emphasizes even spatial dimensions. $\mathfrak{C}^{(2n)}$ is the density of the homotopic invariant associated with background fields, so it is locally a total derivative, $\mathfrak{C}^{(2n)}\equiv\partial_i \mathcal{K}_W^i$. We then perform an integration by part so as to facilitate the implementation of dimensional reduction, i.e.,
\begin{eqnarray}
\phi_W&\equiv &-\text{ch}_W\times \int_{\mathcal{M}^{(2n-1)}} d^{2n-1}\boldsymbol{x} \partial_i\mathcal{I}_W[w(\boldsymbol{x})] \times (\oint_{S^1} d\boldsymbol{x}_{2n}\mathcal{K}^i_W),\label{supp_eq:effective_action}
\end{eqnarray}
where we have assumed that $\mathcal{I}_W[w(\boldsymbol{x})]$ is independent of the $(2n)$-th spatial dimension, and thus the superscript $i\neq 2n$. Then, we assume that $\mathcal{K}_W^i$ is of the Chern-Simons form, i.e., $\mathcal{K}_W^i=\frac{1}{n!}\frac{1}{(2\pi)^{n}}\epsilon^{i i_1 i_2 i_3\dots }A_{i_1}\partial_{i_2}A_{i_3}\dots$, where $A_i$ can be the electromagnetic gauge field, or the gauge field from skyrmions \cite{wilczeck1983prl}. After these preparations, we shall derive the odd dimensional action by integrating out one spatial dimension, implemented by:
\begin{itemize}
\item  Taking the dependence on the extra $2n$-th spatial coordinate dependence exclusively via $A_{2n}$.
\item Inserting half a flux quantum along $\boldsymbol{x}_{2n}$, i.e., $\oint A_{2n}d\boldsymbol{x}_{2n}=-\pi$.
\end{itemize}
Together, we find the following $\phi_W$ for the odd spatial dimension descendant, 
\begin{equation}
\frac{\phi_W}{\pi}=\text{ch}_W \times \int_{\mathcal{M}^{(2n-1)}} \partial_i \mathcal{I}_W[w(\boldsymbol{x})]\times \frac{1}{2\pi}\mathfrak{C}^{(2n-2)} \in \text{ch}_W \times  \mathbb{Z},
\label{supp_eq:action_dr}
\end{equation}
where $\mathfrak{C}^{(2n-2)}$ is independent of $\boldsymbol{x}_i$, and thus $\int d\boldsymbol{x}_i\partial_i\mathcal{I}_W\in 2\pi \mathbb{Z}$ due to spatial periodicity.

Built upon the formula Eq.~\eqref{supp_eq:action_dr}, we now discuss about the ensuing topological order parameter, enumerated below:
\begin{enumerate}
\item For the non-chiral symmetric class, the mixed-state topological order parameter is taken to be $\frac{\phi_W}{\pi}=0,\ 1 \mod 2 \in \mathbb{Z}_2$, which renders a $\mathbb{Z}_2$ invariant. In particular, this $\mathbb{Z}_2$ accounts for the ambiguity encoded in the map between even dimensional parent states and their odd dimensional descendant.
\item For the chiral symmetric class, the descendant state belongs the $\mathbb{Z}$ ($2\mathbb{Z}$) class instead, as chiral symmetry resolves the map ambiguity \cite{ryu2010njp}, which can be detected by $\frac{\phi_W}{\pi} \in\mathbb{Z}$.
\end{enumerate}

\subsection{Example: complex fermion in the one dimensional DIII class}

 \begin{figure}
\includegraphics[scale=0.45]{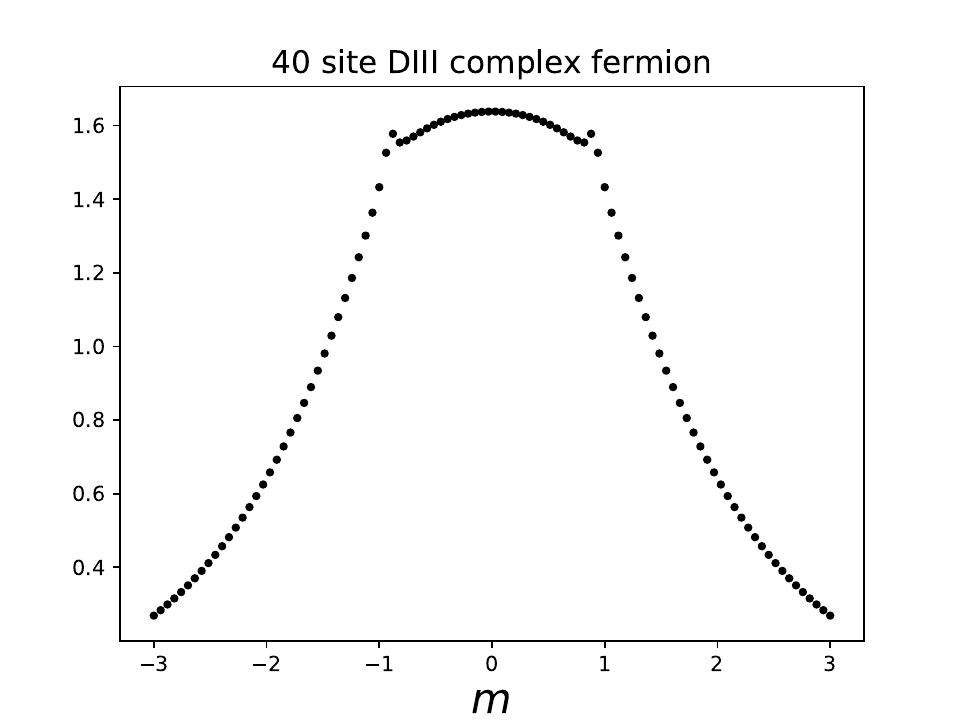}
\includegraphics[scale=0.45]{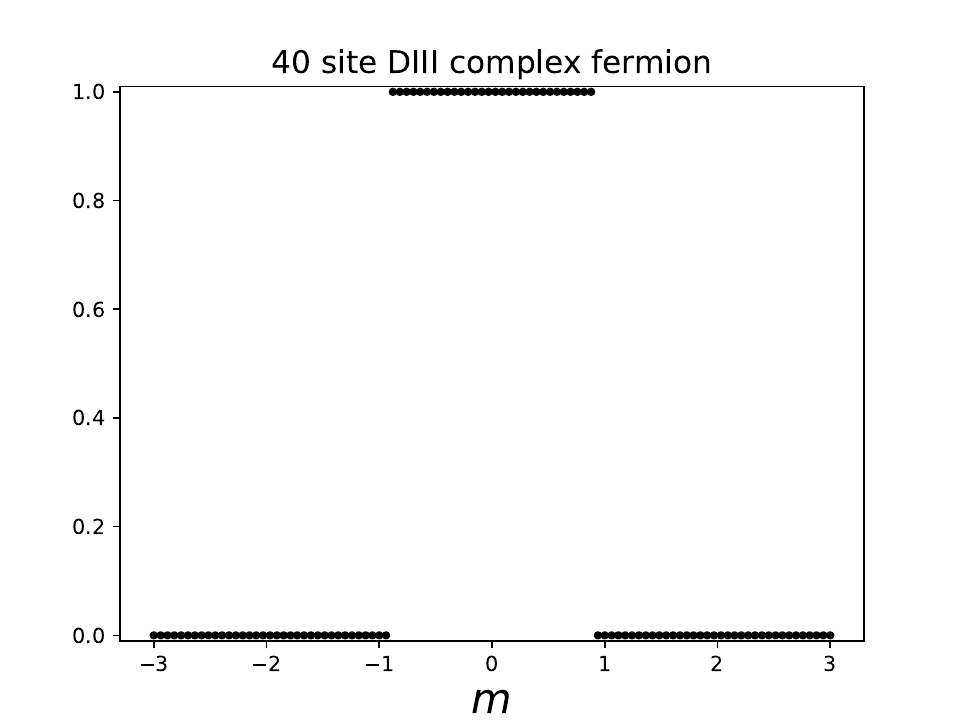}

\caption{Numerical results for the 1D DIII complex fermions, with $G=-\left(\sin k_{x}\sigma^{z}\right)\otimes\tau^{x}+\left(m+\cos k_{x}\right)\sigma^{0}\otimes\tau^{z}$, and $k_x$ for momentum. Here, $\mathcal{W}=\boldsymbol{n}\cdot\boldsymbol{\sigma}\otimes\tau^0$ is the spin matrix, with $\boldsymbol{n}$ a randomly sampled unit vector. Left panel: Plot of $-2\ln|\text{Tr}(\hat\rho \hat{U})|/N_x$ as a function of $m$, with site number $N_x$. Here, we take $\hat{U}=(-1)^{N_x+1} e^{-i\sum_i \hat{\Psi}^\dagger_i \frac{2\pi \boldsymbol{x}}{N_x}\frac{1}{2}(\mathbb{I}+\mathcal{W})\hat\Psi_i}$, and $(-1)^{N_x+1}$ is the factor in reference to the topological trivial phase. This clearly exhibits cusp around the topological transition point (e.g., $m=\pm1$). Right panel: Plot of (modified) $\phi_W$ as a function of $m$, showing a sharp transition between different topological phases.
 \label{supp_fig:numerical_DIII}}
\end{figure}

Based on the strategy outline in the last part, now we construct a mixed state topological order parameter in details, by properly choosing $w[\boldsymbol{x}]$. Following the dimensional reduction sketched in Fig.~\ref{fig:conceptual_plot} in the main text, we shall focus on the second descendant, e.g.,  blue $\mathbb{Z}_2$ in Tab.~\ref{supp_tab:10_class}, while for other classes, one can obtain the mixed-state order parameter via the polarization operator (e.g., generalization of the ensemble geometric phase \cite{bardyn2018prx, huang2022prb}). 

To detect the underlying topology, we shall make a slight modification of the probe operator, according to 
\begin{equation}
\frac{2\pi \boldsymbol{x}_\alpha}{N_i}\mathcal{W}\rightarrow\frac{2\pi \boldsymbol{x}_\alpha}{N_\alpha}\frac{1}{2}(\mathbb{I}+\mathcal{W}),\label{supp_eq:dr}
\end{equation}
where the subscript $\alpha$ refers to the $\alpha$-spatial direction. $\frac{1}{2}(\mathbb{I}+\mathcal{W})$ is a projection operator, activating half the degree of freedom. This modification is necessary because $\text{ch}_W\in 2\mathbb{Z}$ for the $\mathbb{Z}_2$ class (blue in Tab.~\ref{supp_tab:10_class}), inherited from the parent $2\mathbb{Z}$ class  (brown in Tab.~\ref{supp_tab:10_class}). In turn, by taking $w=\frac{2\pi \boldsymbol{x}_\alpha}{N_\alpha}$, $\phi_W\in 2\pi \mathbb{Z}$, and thus $e^{i\phi_W}=+1$. To cure this and improve the resolution of $e^{i\phi_W}$, we utilize Eq.~\eqref{supp_eq:dr} to activate half the degree of freedom. (For superconductors, this modification is not needed, i.e., $e^{i \phi_W} =\pm1$ as explained in Sec. \ref{sec:majorana}.) 


Physically speaking, this modification impacts on both,  the original probe operator (e.g., $\frac{\pi}{N_\alpha}\boldsymbol{x}_\alpha\mathcal{W}$), and the temporal component of the $U(1)$ gauge field (e.g., $A_0= \frac{\pi}{N_\alpha}\boldsymbol{x}_\alpha$, via the overall prefactor $1/2$). It is the modification of the probe operator which provides us with a signal distinguishing different (blue) $\mathbb{Z}_2$ mixed states, i.e., $e^{i\frac{1}{2}\phi_W}=\pm1$, because $w$ associated with $\mathcal{W}$ takes $1/2$ of its original value. The modification of the $U(1)$ part preserves the spatial periodicity, i.e., $\boldsymbol{x}_\alpha\rightarrow\boldsymbol{x}_\alpha+N_\alpha$: Since the $U(1)$ action for (blue) $\mathbb{Z}_2$ is known to be non-topological \cite{qi2010prb, shinsei2012prb1, huang2022prb}, so we expect its contribution to $\phi_W$ vanishes.
Numerically, we confirm this in the DIII complex fermion class, see Fig.~\ref{supp_fig:numerical_DIII}.


\begin{figure}
\includegraphics[scale=0.5]{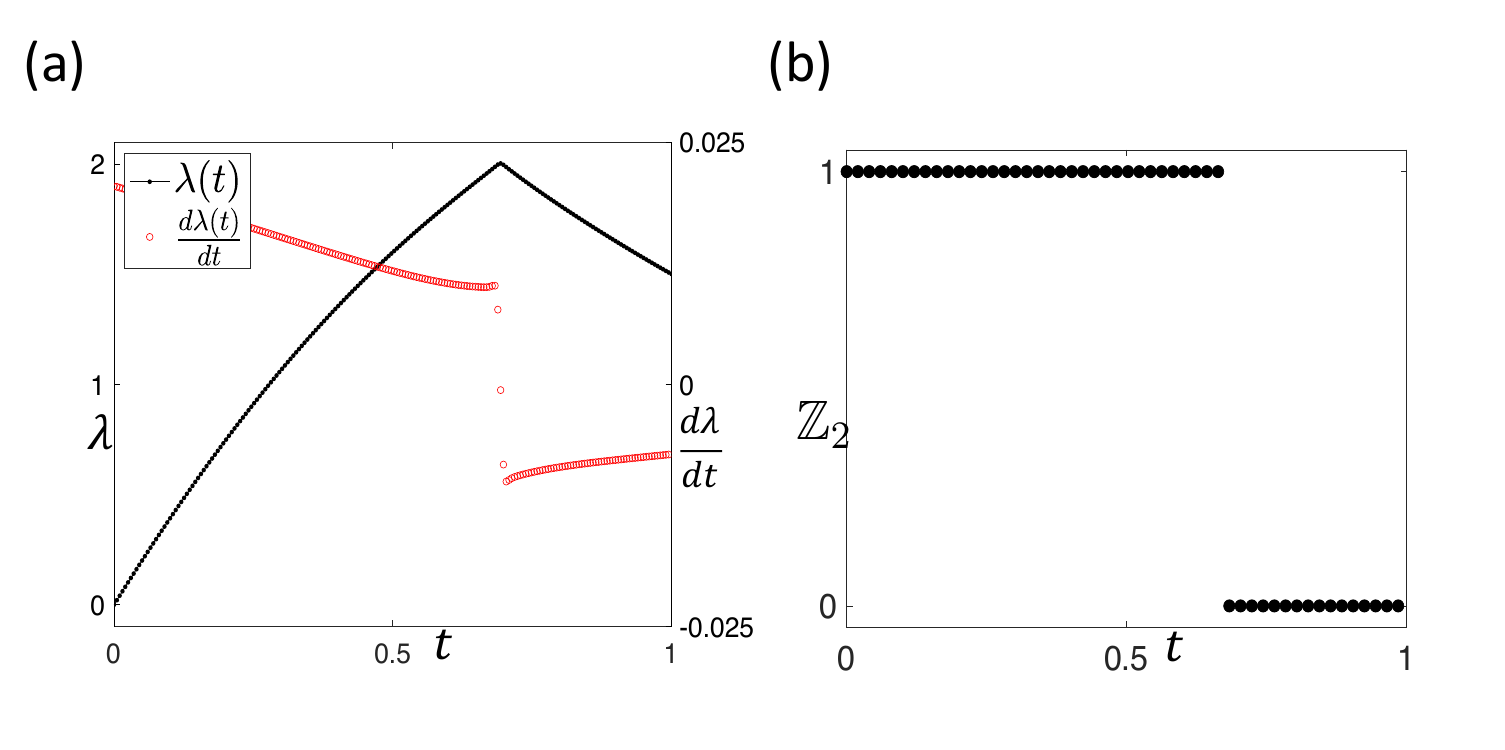}
\caption{Numerical results for mixed state open system evolution starting from a ground state (with one magnetic flux quantum inserted, site number $15\times 15$, $m=1, \Delta=0.5$).  
The left panel shows the amplitude of the Loschmidt echo at $w=\pi$, i.e., $\lambda(t)\equiv-\frac{\ln|\mathcal{Z}_W[w=\pi]|^2}{N_x\times N_y}$, which exhibits a cusp around the transition point. The right panel presents $\frac{1}{2}\frac{\Delta\phi_W}{2\pi}$ as a function of $t$, with gapped $Q_W$.
\label{supp_fig:numerical_BHZ}}
\end{figure}

\section{Numerical results for non-equilibrium mixed state}

We apply our mixed state order parameter to study the non-equilibrium dynamics of mixed states, which arise from exposing the pure ground state of a Hamiltonian to dissipative processes. Concretely, we start from the ground state of the (modified) BHZ model (AII insulator), with Hamiltonian,
\begin{equation}
H=\left(\begin{array}{cc}
H_{0}\left(\boldsymbol{k}\right) & -i\Delta\tau^{y}\\
i\Delta\tau^{y} & H_{0}^{*}\left(-\boldsymbol{k}\right)
\end{array}\right), \ \mathcal{T}=\sigma^{y}\otimes\tau^{0}\mathcal{K}.
\end{equation}
$H_{0}=\boldsymbol{d}\cdot\boldsymbol{\tau}$ with $\boldsymbol{d}=\left(\sin\left(k_{x}\right),\ \sin\left(k_{y}\right),\ m+\cos\left(k_{x}\right)+\cos\left(k_{y}\right)\right)$
is the Qi-Wu-Zhang model \cite{qi2006prb}, and time-reversal symmetry is implemented by $\mathcal T$.

We subsequently examine the ground state subjected to Lindblad dynamics, characterized by linear Lindblad operators (with Hamiltonian switched off)
\begin{equation}
\hat{L}_{i}^{\left(l\right)}=\sqrt{\gamma_{l}}\frac{1+\tau^{z}\otimes\sigma^{0}}{2}\hat{\psi}_{i},\ \hat{L}_{i}^{\left(g\right)}=\sqrt{\gamma_{g}}\frac{1-\tau^{z}\otimes\sigma^{0}}{2}\hat{\psi}_{i}^{\dagger},
\end{equation}
which preserves the time-reversal symmetry \citep{altland2021prx, mao2023arxiv}. These processes by themselves target a trivial pure steady state, i.e., $|\psi_{\text{target}}\rangle=\prod_{i}\hat{\psi}^\dagger_{i,\ a=2}\hat{\psi}^\dagger_{i,\ a=4}|0\rangle$, where $|0\rangle$ is for the vacuum without particle, and the subscript $a$ for internal indices.

For the probe operator, possible choices for the $\mathcal{W}$ matrix are then $\mathcal{W}=\{\boldsymbol{\sigma}\otimes\tau^{x,\ z,\ 0},\ \sigma^{0}\otimes\tau^{y}\}$. 
Numerical results are presented in Fig.~\ref{supp_fig:numerical_BHZ}, which include the plot of the Loschmidt echo $|\mathcal{Z}_W[w=\pi]|^2$ at $w=\pi$, or its rate function $\lambda(t)\equiv -\frac{\ln |\mathcal{Z}_W[w=\pi]|^2}{N_x \times N_y}$] in the left panel, and mixed state order parameter in the right panel. Here, both the rate function $\lambda(t)$, and $\frac{\Delta\phi_W}{2\pi}$ provide a sharp signature distinguishing different phases.

\section{Phase signal $\phi_W$ in cold atomic ensembles}

\begin{figure}
\includegraphics[scale=0.5]{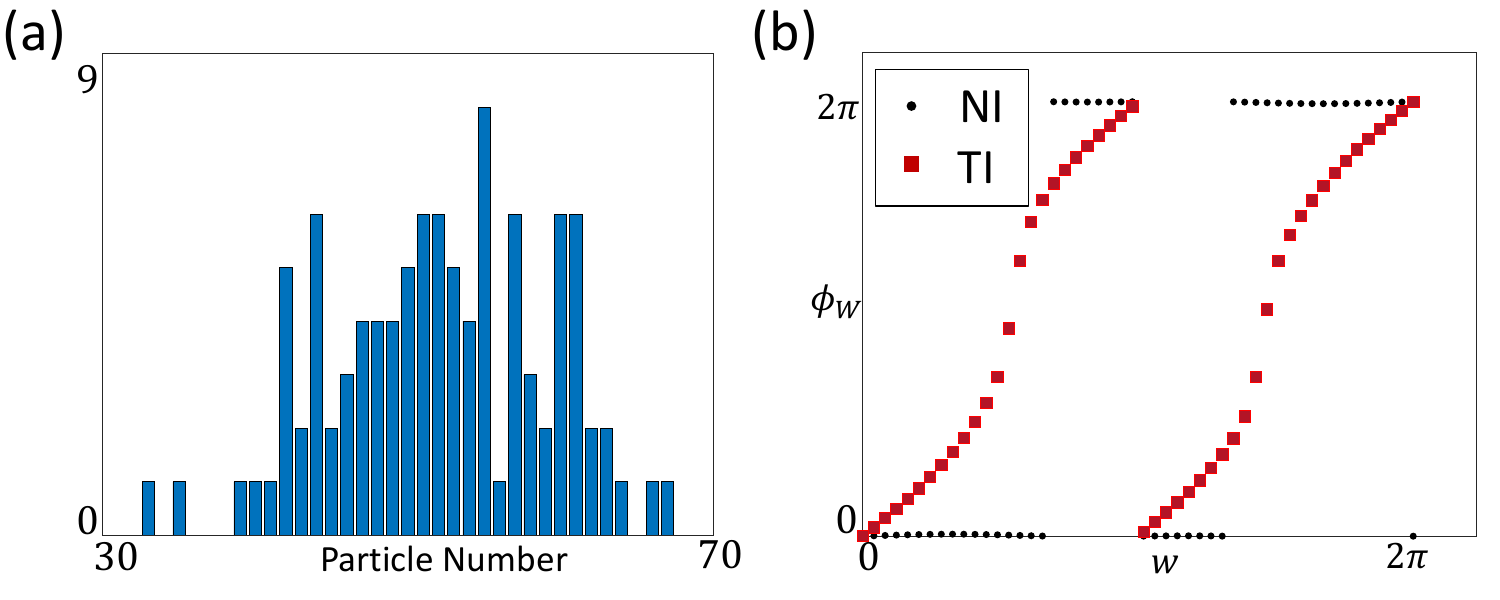}
\caption{Numerical results for the phase signal $\phi_W$ in cold atomic ensembles,  focusing on the modified BHZ model on a $5\times 5$ lattice; each lattice site hosts 4 states. The ensemble average particle number corresponding to half filling is $50$, and we  set the model parameters to $\beta=3$ and $\Delta =0.5$. In (a), we have sampled $50$ runs from the Poisson distribution, where the $y$-axis represents the frequency of particle number. In (b), we delineate the ensemble averaged phase signal for both topological insulators (TI), and normal insulators (NI), exemplified by $m=1, 3$, respectively. The ensemble averaged phase signal for $N_s$ samples is defined as $\phi_W\equiv\text{arg}[\frac{1}{N_s}\sum_{n=1}^{N_s} \mathcal{Z}_{W}^{(n)}]$, with $n$ the sample index.
\label{supp_fig:ensemble_mBHZ}}
\end{figure}

In the main text, we have discussed numerical results for $\phi_W$ within the canonical ensemble, i.e., for a fixed number of particles. Here we demonstrate the effectiveness of the order parameter $\frac{\Delta\phi_W}{2\pi}$ in full counting statistic measurements with cold atoms. These measurements will be carried out by repeating the counting experiment many times, with two key characteristics: (1) Within each run, the particle number is fixed, and (2) For every run however, the particle number might be different, governed by some probability distribution function for the particle number. In Fig. ~\ref{supp_fig:ensemble_mBHZ}, we present numerical results incorporating these specifics for the modified BHZ model (Eq.~\eqref{eq:mbhz} in the main text), adopting a Poisson distribution function to model the mean particle number at half-filling for definiteness (see Fig.~\ref{supp_fig:ensemble_mBHZ} (a)). The results depicted in Fig.~\ref{supp_fig:ensemble_mBHZ} (b) clearly demonstrate that $\frac{\Delta\phi_W}{2\pi}$ efficiently differentiates between topological and  trivial phases. 

\section{Results for superconductors (Majorana fermions) }\label{sec:majorana}
We present results for superconductors (Majorana fermions), which includes (i) the relation between spectral asymmetry and fermion parity, and (ii) the construction of a mixed state topological order parameter. 

We first introduce the necessary notation. In the context of superconductors, the $U(1)$ symmetry is broken down to the discrete $\mathbb{Z}_2$ fermion parity symmetry. Accordingly, we work in the Nambu basis, and the modular Hamiltonian is expressed as
\begin{equation}
\hat{G} = \hat{\Psi}^\dagger G\hat{\Psi}, \ \text{with} \ \hat{\Psi}\equiv (\hat\psi,\ \hat\psi^\dagger)^T,
\end{equation}
where the first quantized operator $G$ embodies particle-hole symmetry.

\subsection{Fermion parity and spectral asymmetry for Majorana}
We demonstrate that fermion parity detects spectral properties of the modular Hamiltonian $G$. 
Namely, due to particle-hole symmetry, the  eigenvalues of the matrix $G$ appear in pairs with opposite sign, i.e., $\left\{ \lambda_{n},\ -\lambda_{n}\right\}$, where $\lambda_{n}$ can be positive or negative. We then divide these particle-hole pairs to two different classes, e.g., $\left\{ \lambda_{1},\ \lambda_{2},\ \dots\right\}$  and $\left\{ -\lambda_{1},\ -\lambda_{2},\ \dots\right\}$ , such that $G=\oplus_{n}\left(\begin{array}{cc}
0 & i\lambda_{n}\\
-i\lambda_{n} & 0
\end{array}\right).$
It turns out that the fermion parity  $\left(-1\right)^{\hat{Q}}$ crucially hinges on the spectral asymmetry of $\left\{ \lambda_{1},\ \lambda_{2},\ \dots\right\}$ , 
\begin{equation}
\text{Im}\ln\text{Tr}\left[\left(-1\right)^{\hat{Q}}\hat{\rho}\right]=-\frac{\pi}{2}N+\frac{\pi}{2}\sum_{n}\text{sign}\left(\lambda_{n}\right),\label{supp_eq:spec_majorana}
\end{equation}
where $N$ is for the total particle number, and we have used the following identity (see, for example, \cite{klich2014jsm, grabsch2019adp})
\begin{equation}
\text{Tr}\left(\hat{\rho}e^{-i\pi\hat{Q}}\right)=\left(-i\right)^{N}\text{Pf}\left(\tanh G\right).\label{supp:fermion_parity}
\end{equation}
This indicates that the fermion parity probes the asymmetry of $\lambda_n$.


\subsection{Probe operator for superconductors in the prime series}
We now turn to Majorana models in the prime series, which includes examples like the Kitaev chain, and the 2D p-wave chiral superconductor. The superconductor no longer enjoys $U(1)$ symmetry, but still we can induce a spectral asymmetry via a $\mathbb{Z}_2$ gauge field, implemented via anti-periodic spatial boundary conditions, based upon which we construct a mixed state order parameter. 

 The probe operator, similar to the complex fermion counterpart, is taken to be
\begin{equation}
e^{-i w \hat{Q}},\ \hat Q=\sum_{i,\ a}\hat\psi^\dagger_{i,\ a} \psi_{i, a},
\end{equation}
For the mixed state density matrix, we need to replace the $U(1)$ gauge fields by and alternative construction. In fact, $\mathbb{Z}_2$ gauge fields can be utilized to activate topological charge. This can be implemented in the density matrix by twisted boundary conditions according to
\begin{equation}
\hat{\rho}=e^{-\hat{G}}|_{(s_1, s_2, \dots)}, 
\end{equation}
where $s_1, s_2\dots =0, \ \pi$ for the periodic/antiperiodic boundary condition along $i$-th spatial direction. The ensuing mixed state order parameter is then defined as the winding number of the following phase
\begin{equation}
\phi_W(w)\equiv \text{arg}[\prod_{s_1, s_2, \dots =0, \pi}\text{Tr}(\hat\rho e^{-i w \hat Q})^{\nu_{\{ s \}}}],
\end{equation}
where $\nu_{\{s\}}=+1 (-1)$ for $\{s\}= (s_1, s_2, ...)$ containing even (odd) number of $0$. Here, the  product provides a normalization which ensures that the winding number $\phi_W(w)$ is only from insertion of $\mathbb{Z}_2$ gauge field in all spatial directions, For example, the factor $-\frac{\pi}{2}N$ in Eq.~\eqref{supp_eq:spec_majorana} is canceled, which shall be further illustrated below, via examples.

We are now in the position to construct the associated effective action, from which we can infer the descendant $\mathbb{Z}_2$ invariant, via the method of dimensional reduction. That is, the action is
\begin{equation}
\phi_W(w)=\text{ch}_W\int \mathcal{I}_W[w]\frac{a_1}{\pi}\times \frac{a_2}{\pi}\dots,\ \text{with}\ \text{ch}_W\in\mathbb{Z}, \ \mathcal{I}_W[w]|_{w=0}^{w=2\pi}=2\pi,\  \text{and}\ \mathcal{I}_W[w=\pi]=\pi,
\end{equation}
where $a_i$ is the $\mathbb{Z}_2$ gauge field for the $i$-th direction, such that $\oint a_i =s_i$. We have assumed that $\mathcal{I}_W[w]$ is a smooth function of $w$ for a gapped system. 
Also, $\mathcal{I}_W[w=\pi]=\pi$ is from Hermiticity, i.e., $\phi_W(w)=-\phi_W(-w)$. Taking one spatial dimension as an example, this reproduces the result in Ref.~\cite{karch2019scipost}, by taking $\mathcal{I}_W[w]=w$. Finally, it is worth mentioning that the effective action shares the same form in all dimensions, as the dimensional reduction is implemented by integrating out a $\pi$ flux inserted along the extra dimension. 

As an illustration, we consider the $w=\pi$ point, which determines the eveness/oddness of the winding number (i.e., $\text{ch}_W$). In the one-dimensional case, we find 
\begin{equation}
e^{i\phi_W(\pi)}=\text{sign}\left[\frac{\langle(-1)^{\hat{Q}}\rangle|_{s_1=\pi}}{\langle(-1)^{\hat{Q}}\rangle|_{s_1=0}}\right] \label{supp:fermion_parity_pf},
\end{equation}
 which reproduces the $\mathbb{Z}_2$ index for Majorana fermions \cite{kitaev2001pu, karch2019scipost}. We see the role of the denominator,  canceling out irrelevant factors, e.g., $(-i)^N$ in Eq.~\eqref{supp:fermion_parity}.  Meanwhile, the two-dimensional counterpart is 
 \begin{equation}
e^{i\phi_W(\pi)}=\text{sign}\left[\frac{\langle(-1)^{\hat{Q}}\rangle|_{s_1=\pi, s_2=\pi}\times \langle(-1)^{\hat{Q}}\rangle|_{s_1=0, s_2=0}}{\langle(-1)^{\hat{Q}}\rangle|_{s_1=0, s_2=\pi}\times \langle(-1)^{\hat{Q}}\rangle|_{s_1=\pi, s_2=0}}\right],
\end{equation}
which is reminiscent of the invariant defined in Ref.~\cite{ghosh2010prb}.
Physically, this describes the pumping of the one-dimensional fermion parity [e.g., Eq.~\eqref{supp:fermion_parity_pf}], after insertion of a $\pi$ flux (i.e., $s_2=\pi$).

 \subsubsection{Numerical results for $\phi_W$ in one- and two- dimensions}

 \begin{figure}
\includegraphics[scale=0.35]{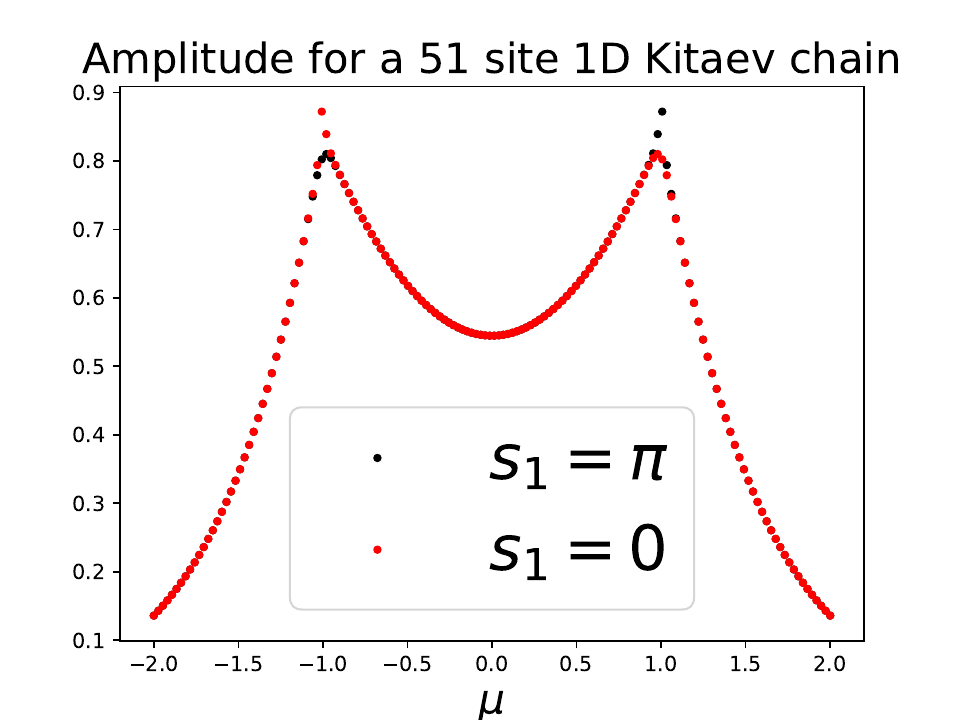}
\includegraphics[scale=0.35]{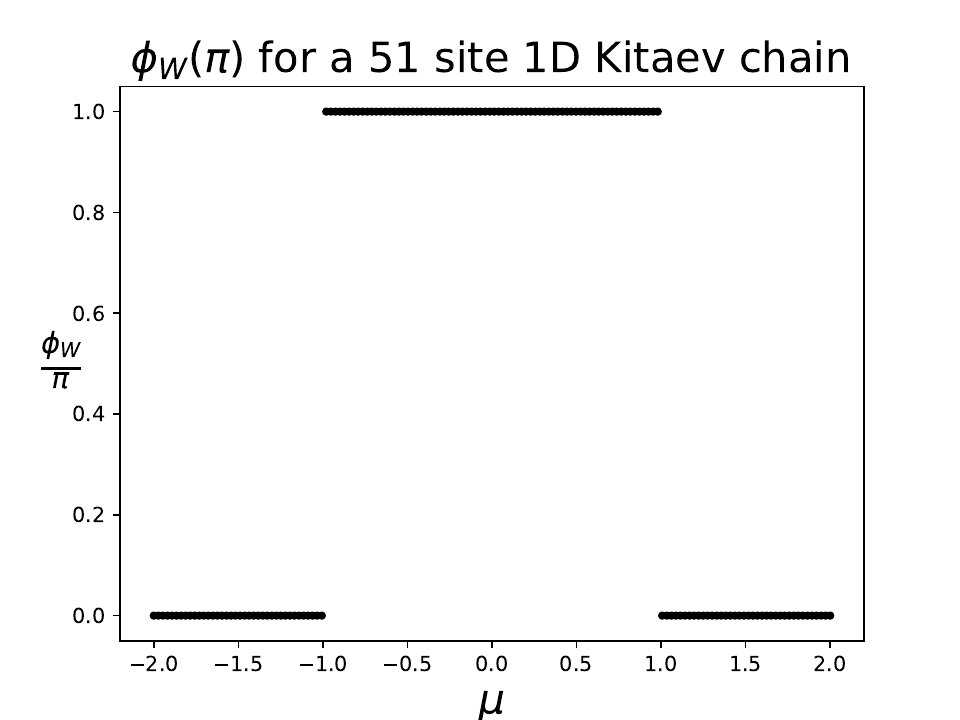}
\includegraphics[scale=0.35]{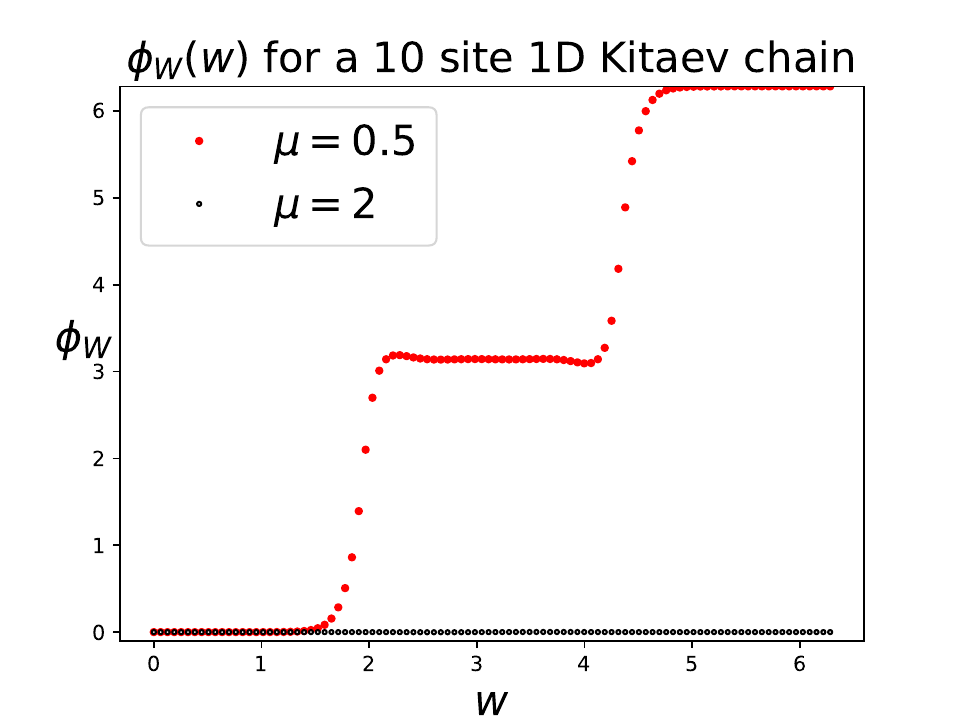}

\caption{Numerical results for the 1D Kitaev chain. Left panel: Plot of $-2\ln|\text{Tr}[\hat\rho (-1)^{\hat Q }]|_{s_1=0, \pi}|/N_x$ as a function of $\mu$, which  exhibits singular points around the topological transition point ($\mu=\pm1$). Middle panel: Plot of $\frac{\phi_W(w=\pi)}{\pi}$ as a function of $\mu$, which reconstructs the zero-temperature phase diagram. Right panel: Plot of $\phi_W(w)$ at $\mu=0.5, 2$ as a function of $w$, which shows that $\phi_W$ possesses different winding numbers in the topologically trivial and non-trivial phase.
 \label{supp_fig:numerical_kitaev}}
\end{figure}

 To further support our results above, we shall present numerical results in the Kitaev chain, and the two dimenisonal chiral p-wave superconductors. 
 
For the Kitaev chain, the  modular Hamiltonian is 
 \begin{eqnarray}
\hat{G}&=&-\mu \sum_i (\hat{\psi}_i^\dagger \hat{\psi}_i+h.c.) -t\sum_i (\hat{\psi}^\dagger_{i+1}\hat{\psi}_i+h.c.)+\Delta \sum_i(\hat{\psi}_{i+1} \hat{\psi}_{i}+h.c.).
\end{eqnarray}
For later convenience, we shall set $t=\Delta=1$, so this model is topologically nontrivial (trivial) for $|\mu|<1$ ($|\mu|>1$). Meanwhile, the twisted spatial boundary condition is implemented via 
\begin{equation}
-t e^{i s_1}\hat{\psi}_1^\dagger \hat\psi_{N_x},\ \text{and}\ \Delta e^{i s_1}\hat\psi_{1}\hat\psi_{N_x},\ \text{with}\ s_1=0,\ \pi.
\end{equation}
Numerical results for the Kitaev chain are presented in Fig.~\ref{supp_fig:numerical_kitaev}. The left panel and the middle panel demonstrate that $\frac{\phi_W(w=\pi)}{\pi}$ successfully captures the topology in different phase, and the transition point manifests itself as a cusp in the amplitude, $\lambda \equiv -\frac{\ln|\mathcal{Z}_W|^2}{N_x}$. The right panel displays $\phi_W$ as a function of $w$, which does exhibit different winding for the topologically trivial and non-trivial phases. Note that the trace formula for the product of two Majorana Gaussian operators is not known except when one of the operator is the fermion parity (to the best of our knowledge) \cite{klich2014jsm}, so here we used the method of exact diagonalization.

 \begin{figure}
\includegraphics[scale=0.45]{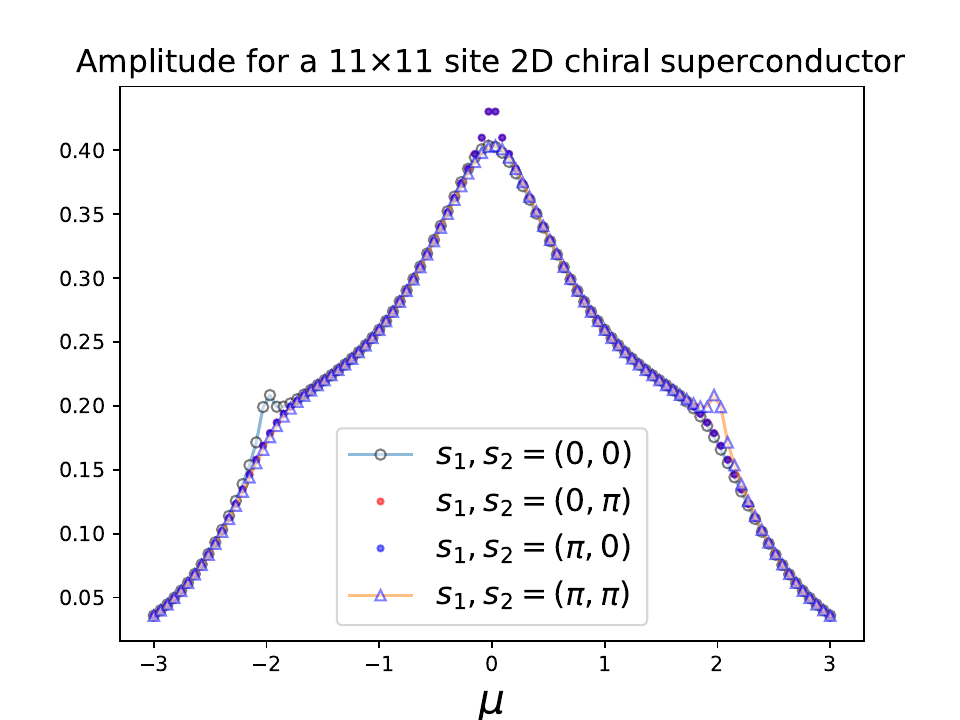}
\includegraphics[scale=0.45]{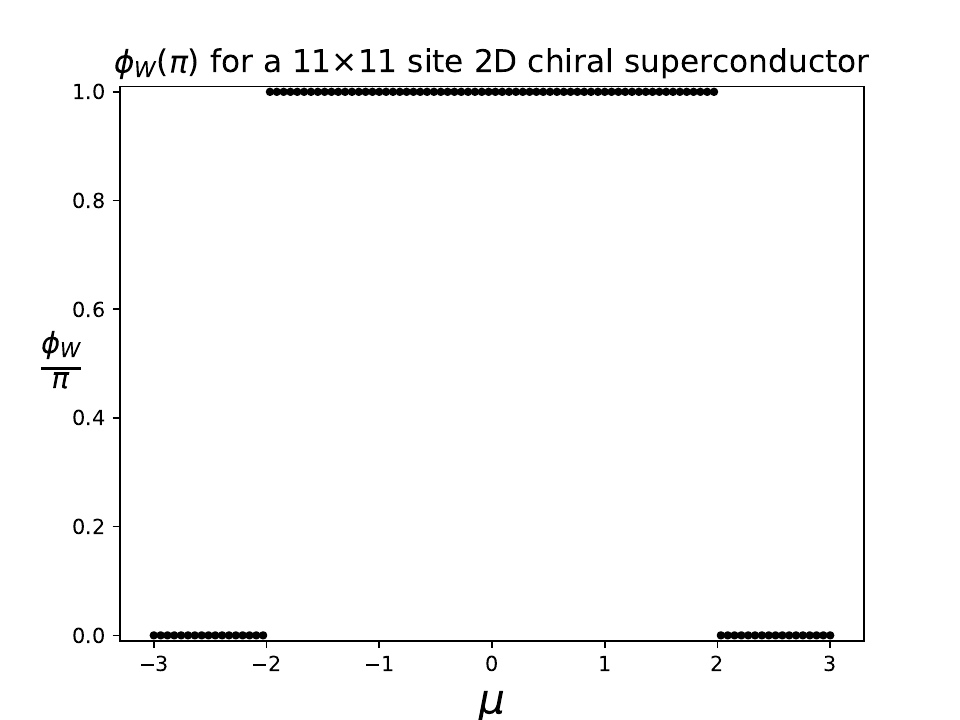}

\caption{Numerical results for the 2D chiral superconductor. Left panel: Plot of $-2\ln|\text{Tr}[\hat\rho (-1)^{\hat Q }]|_{s_1, s_2 =0, \pi}|/(N_x \times N_y)$ as a function of $\mu$, which  exhibits a singular point around the topological phase transition point ($\mu=\pm 2$). Right panel: Plot of $\frac{\phi_W(w=\pi)}{\pi}$ as a function of $\mu$, which shows that $\frac{\phi_W(w=\pi)}{\pi}$ detects the topologically trivial and non-trivial phases.
 \label{supp_fig:numerical_2dchiral_sc}}
\end{figure}

For the two-dimensional chiral p-wave superconductor, the  modular Hamiltonian is 
\begin{eqnarray}
\hat{G}&=&\sum_{m, n}[-t (\hat{\psi}_{m+1, n}^\dagger \hat{\psi}_{m, n}+h.c.)-t(\hat{\psi}_{m, n+1}^\dagger \hat{\psi}_{m, n}+h.c.)]\nonumber\\
{}&&-\sum_{m, n}\mu(\hat{\psi}^\dagger_{m, n}\hat{\psi}_{m, n}+h.c.)\nonumber\\
{}&&+\sum_{m, n}[(\Delta \hat{\psi}^\dagger_{m+1, n} \hat{\psi}^\dagger_{m, n}+h.c. )+(i\Delta \hat{\psi}^\dagger_{m, n+1}\hat{\psi}^\dagger_{m, n}+h.c.)],
\end{eqnarray}
and the twisted boundary condition is implemented similar to the Kitaev chain. Also, for later convenience, we set $\Delta=t=1$, and this model is  topologically nontrivial (trivial) for $|\mu|<2$ ($|\mu|>2)$. Numerical results are presented in Fig.~\ref{supp_fig:numerical_2dchiral_sc}, which confirms that $\frac{\phi_W(w=\pi)}{\pi}$ probes the underlying topology for mixed states.

\subsection{Probe operator for superconductors in the $\mathbb{Z}_2$ class}
For superconductors in the $\mathbb{Z}_2$ class, parallel to their complex fermion counterparts, we are required to choose $\mathcal{W}$ differently from $\mathbb{I}$ in order to resolve the topological signal. In turn, constraints for $\mathcal{W}$ are similar, except: 
\begin{enumerate}
\item $\mathcal{W}$ should be antisymmetric, due to the anti-commuting nature of Majorana; \label{supp_item1_Z2_sc} 
\item The $\mathcal{W}\rightarrow \frac{1}{2}\mathcal{W}$ with a $\frac{1}{2}$ factor, from the redudancy in the Nambu space.\label{supp_item2_Z2_sc}  
\end{enumerate}

\
This construction hinges on the following identity for Majorana fermions, 
\begin{equation}
[\text{Tr}\left(e^{-\hat{\gamma}A\hat{\gamma}}e^{-\hat{\gamma}B\hat{\gamma}}\right)]^2=\det(\mathbb{I}+e^{-2A}e^{-2B}),
\end{equation}
where the anticoummtation relation of $\hat\gamma$ is $\{\hat{\gamma}_a,\ \hat{\gamma}_b\}=\delta_{ab}$. $A,\ B$ are skew symmetric, from the anti-commutation relations, reproducing \ref{supp_item1_Z2_sc}. The Majorana Gaussian operator $e^{-\hat{\gamma} A \hat{\gamma}}$ appears on the right-hand side as $e^{-\boldsymbol{2}A}$, leading to \ref{supp_item2_Z2_sc}.



\begin{thebibliography}{97}
\bibitem [{FN_()}]{FN_od}%
  \BibitemOpen
  \href@noop {} {}\bibinfo {note} {Complex fermions in odd spatial dimension
  belonging to the second descendant states, requires the following
  modification: $\mathcal{W}\rightarrow \frac{1}{2}(\mathbb{I}+\mathcal{W})$,
  which singles out the eigenvectors with positive eigenvalues. This stems from
  $e^{i\phi_W} = +1$ for $\mathcal{W}$, reflecting an even Chern number for its
  parent state. Instead, ultilizing $\frac{1}{2}(\mathbb{I}+\mathcal{W})$
  results in $e^{i\phi_W} = \pm 1$, revealing the underlying topology
  \cite{supp}. For superconductors in the same class, such modifcation is not
  needed, since they contain only half the degree of freedom compared to their
  complex fermion counterpart.}\BibitemShut {Stop}%
\bibitem [{wtr()}]{wtrace_fn}%
  \BibitemOpen
  \href@noop {} {}\bibinfo {note} {We have neglected a term
  $\text{tr}\mathcal{W}$. This is because: 1. For $\mathcal{W}=\mathbb{I}$, it
  gives a state-independent phase factor; 2. In the presence of time-reversal
  or particle-hole symmetry, it vanishes. Let us make this explicit for time
  reversal symmetry: \ref{list:2a} and \ref{list:2b} imply that
  $\text{Tr}\mathcal{W}=-\text{Tr}\mathcal{W}=0$.}\BibitemShut {Stop}%
\bibitem [{fig()}]{fig2_fn}%
  \BibitemOpen
  \href@noop {} {}\bibinfo {note} {We keep results obtained from gaped $Q_W$ so
  as to ensure $\phi_W$ well-defined.}\BibitemShut {Stop}%
\end{thebibliography}
\end{document}